\documentclass[12pt,preprint]{aastex}

\newcommand{\hii}{H\,II}
\newcommand{\w}{\,$\lambda$\,}
\newcommand{\ww}{\,$\lambda\lambda$\,}
\newcommand{\teff}{$T_{\!\mbox{\scriptsize \em eff}}$}
\newcommand{\zsun}{$Z_\odot$}
\newcommand{\msun}{$M_\odot$}
\newcommand{\ebv}{$E(B-V)$}
\newcommand{\ew}{$W(H\beta)$}

\shorttitle{Metal-rich H\,II regions}
\shortauthors{Bresolin \& Kennicutt}

\received{2001 December 13}
\begin{document}

\title{Optical spectroscopy of metal-rich H\,II
regions and circumnuclear hot spots in M83 and NGC~3351}

\author{Fabio Bresolin}
\affil{Institute for Astronomy, 2680 Woodlawn Drive, Honolulu HI
96822}
\email{bresolin@ifa.hawaii.edu}

\and

\author{Robert C. Kennicutt, Jr.}
\affil{Steward Observatory, University of Arizona, Tucson AZ 65721}
\email{robk@as.arizona.edu}

\begin{abstract}
We present optical spectroscopy of a sample of metal-rich
extragalactic \hii\/ regions in the spiral galaxies M83 (NGC~5236),
NGC~3351 and NGC~6384. The metal abundance, estimated from
semi-empirical methods using strong emission lines, is found to be
above solar for most of the objects. The sample includes a number of
circumnuclear
\hii\/ regions (hot spots), which are found in general to have
spectral properties similar to the \hii\/ regions located in the
disks. Different age estimators (equivalent width of the $H\beta$
emission line \ew, Balmer line absorption profiles, UV spectra)
provide consistently young ages (4-6 Myr) for the hot spots.

We have detected the W-R bump feature at 4650~\AA\/ in five of the
objects, and in fewer cases possibly the WC features in the red at
5696~\AA\/ and 5808~\AA. Six additional objects showing W-R features
are drawn from our previous work on extragalactic \hii\/ regions.
From the measured luminosity of the stellar He\,II\w4686~\AA\/ line we
estimate a small number of WN stars (from 1-2 to about 30).  We have
compared the measured intensities and equivalent widths of the W-R
bump to the predictions of recent evolutionary models for massive
stellar populations. By assuming instantaneous bursts of star
formation the ages derived from \ew\/, between 3 and 6 Myr, are in
agreement with the age span predicted by the models. The observed
strength of the W-R bump is in agreement with the predictions for only
half of the objects, the remainder showing lower I(4650)/I($H\beta$)
and $W(4650)$ than the models of the appropriate metallicities. Our
favored explanation is related to the finite number of stars formed in
the clusters and to stochastic effects likely to be at work in star
forming regions of the size considered here.

The He\,I\,\w5876 recombination line is used as an indicator of the
equivalent effective temperature of the ionizing clusters, \teff. We
have found that this temperature is not affected by the presence of
W-R stars. From a comparison with published photoionization models
based on synthetic cluster spectral energy distributions we find some
evidence for an overestimation of the number of He ionizing photons in
the model fluxes. In general, however, the massive star diagnostics
considered in our work are in agreement with the predictions of recent
evolutionary models calculated with a Salpeter initial mass function
and a high upper mass limit.  We find no compelling evidence for a
depletion of massive stars ($M$\,$>$\,40-50~\msun) in the initial mass
function of metal-rich clusters, contrary to our previous conclusions
based on older evolutionary models.

\end{abstract}


\section{Introduction}

During the course of the last two decades the vast majority of
Extragalactic \hii\/ Region (EHR) studies have been concentrated on
high-excitation, low-metallicity objects ($Z$\,$\leq$\,0.5\,\zsun,
where we adopt $12+\log (O/H)_\odot$\,=\,8.9).  Aside from the
scientific drivers of different authors mostly interested in
metal-poor nebulae (high-redshift galaxies, primordial helium
abundance, \hii\/ galaxies), this is in part also due to a selection
effect (decreasing luminosities and emission-line equivalent widths
with increasing metallicity), as well as to intrinsic difficulties in
the investigation of such objects, which are related to their physical
status.  The auroral lines used as electron temperature ($T_e$)
diagnostics (mostly [O\,III]\,$\lambda4363$, but also
[N\,II]\,$\lambda5755$ and [S\,III]\,$\lambda6312$), and necessary for
nebular abundance determinations, are routinely observed at low
metallicity, but become fainter and eventually unobservable with
increasing metallicity.  The enhanced cooling of the gas from the
metals, mainly through oxygen forbidden lines, is responsible for low
equilibrium $T_e$ values and correspondingly weak optical [O\,III]
lines. Therefore, `strong line' semi-empirical methods for abundance
measurement at high metallicity have been developed, the most widely
used being that proposed by \citet{pagel79}:
$R_{23}$\,=\,([O\,II]\,$\lambda3727$~+~[O\,III]\,$\lambda\lambda$4959,5007)/H$\beta$.
While at low metallicity these empirical abundance indicators can be
directly calibrated with the available $T_e$ measurements
(\citealt{skillman89}, \citealt{pyliugin00}), at high metallicity
($12+\log(O/H)>8.6$) their calibration must rely on nebular
photoionization models, but different approaches may lead to
abundances differing by factors of two in the supersolar regime
(\citealt{ep84}, \citealt{mccall85}, \citealt{dopita86},
\citealt{mcgaugh91}, \citealt{pyliugin01}).
Given the fact that high-metallicity \hii\/ regions comprise the
majority of star forming regions in early-type spiral galaxies and in
the central regions of most late-type spirals
(\citealt{vilacostas92}), this has important consequences for the
study of abundance gradients and their interpretation in terms of
models of galactic chemical evolution.  A further complication arises
from the fact that even when weak auroral lines are detected in
metal-rich nebulae, their strength cannot be used directly to
determine $T_e$, due to the effect of temperature inhomogeneities
within the nebulae, which would lead to underestimate the abundance by
large factors using standard techniques (\citealt{garnett92},
\citealt{stasinska01}). Nevertheless, the detection of these faint
lines at such metallicities provides additional constraints for
photoionization models in which one attempts to reproduce the observed
line intensity ratios as a function of the nebular parameters, and in
particular of the element abundances.

In recent years the interest regarding metal-rich EHRs has been
revived, mainly for two reasons. First, the detection of very weak
auroral lines in a handful of objects has led to the determination of
abundances for nebulae of solar-like metallicity, e.g.~\/Searle~5 in
M101 ($12+\log (O/H)=8.9$, \citealt{kinkel94}), and CDT1 in NGC~1232
($12+\log (O/H)=8.95$, \citealt{castellanos01}). These studies suggest
a revision of the empirical abundance calibrators currently employed,
and in the case of $R_{23}$ making the derived $O/H$ abundances
smaller by approximately a factor of two at $Z>\;$\zsun\/ (see also
\citealt{diaz00}).

Secondly, the stellar content of high-metallicity EHRs might have
different properties than those at low metallicity. This issue is
related to the more general question of possible Initial Mass Function
(IMF) variations in starbursts. Although there is mounting evidence
that most high-excitation EHRs and \hii\/ galaxies are ionized by
star clusters following a Salpeter slope IMF with a high upper mass
cutoff ($\sim$100 M$_\odot$; \citealt{stasinska96},
\citealt{masseyhunter98}, \citealt{massey98}, \citealt{tremonti01}, 
\citealt{delgado01}), arguments for differing IMF slopes or mass 
cutoffs in starburst galaxies have been presented (\citealt{luhman98},
\citealt{achtermann95}). 
In the case of metal-rich EHRs, \citet[hereafter BKG]{bresolin99} have
argued that the upper cutoff mass of the (Salpeter slope) IMF of the
ionizing clusters might be as low as 30~$M_\odot$. Using the
He\,I\,\w5876 line as a stellar thermometer, sensitive to the
30,000$\;<\;$\teff$\;<\;$40,000~K range of the ionizing cluster
effective temperature, they found that the empirical decrease in
\teff\/ with increasing metallicity could not be accounted for simply 
by metallicity effects on stellar evolution models
(\citealt{maeder90}), and might be related to the absence of hot,
massive stars at above-solar metallicities. Although upper mass
cutoffs as low as 30 $M_\odot$ seem to be ruled out by more recent
modeling of optical and mid-infrared spectroscopy of starbursts
(\citealt{schaerer00},
\citealt{thornley00}), and massive stars are observed in the
metal-rich Galactic center (\citealt{figer98}), there are additional
indications of a decrease in the upper mass cutoffs in starbursts at
high metallicity (e.g.~\/\citealt{goldader97},
\citealt{coziol01}).
Some discrepancies between the observed line strength ratios and the
theoretical values also led BKG to suggest that the predictions of
current stellar models, in particular those regarding the ionizing
output during the Wolf-Rayet (W-R) phase, might need some revision
(see also \citealt{crowther99}, \citealt{smith02}), which could make low
upper cutoff masses in the IMF unnecessary to explain the
observations.

To address the question of possible peculiarities of the EHR stellar
content at high metallicity, we have obtained new spectrophotometry of
circumnuclear \hii\/ regions, or `hot spots' (\citealt{morgan58},
\citealt{sersic65,sersic67}) in the spiral galaxies
M83 and NGC~3351, where the oxygen abundance is expected to be high,
due to the small galactocentric distances. Additionally, EHRs located
in the inner disk of these galaxies, as well as in NGC~6384, were also
observed.  These are also expected to be metal-rich, as indicated by
studies of extragalactic abundance gradients (\citealt{dufour80},
\citealt{vilacostas92},
\citealt{oey93}, \citealt{dutil99}). While most of the recent studies of
circumnuclear \hii\/ regions have been carried out in the
near-infrared, optical spectroscopy is still valuable to compare the
properties derived for the hot spots with the vast amount of optical
emission line studies of normal EHRs located in the disks of spirals
(\citealt{mccall85}, \citealt{vilacostas92},
\citealt{zaritsky94}, BKG).

With its intense star forming activity, the nuclear region of M83
(NGC~5236; $D$\,=\,3.2 Mpc, assuming the same distance as the
companion galaxy NGC~5253, \citealt{freedman01}) is one of the nearest
examples of the circumnuclear starburst class of objects, which
includes other recently studied nuclear hot spot galaxies, e.g.~\/M100
(\citealt{ryder01}), NGC~2903 (\citealt{herrero01}), NGC~1512 and
NGC~5248 (\citealt{maoz01}).  The current understanding of
star-forming nuclear rings is that they lie near the inner Lindblad
resonance of a barred galaxy, where the bar-driven gas inflow causes
the gas to accumulate, triggering an intense episode of star formation
(\citealt{buta93}, \citealt{elmegreen94},
\citealt{shlosman99}).
The overall structure of the nuclear ring in M83 emerging from
spatially detailed studies at different wavelengths, mainly in the IR
where the effects of dust extinction are reduced, is rather complex,
with a string of hot spots distributed along an arc extending 7 arcsec
(110 pc) SW of the nucleus (\citealt{gallais91},
\citealt{elmegreen98}, \citealt{thatte00}).
Within the central 300 pc, containing $\sim3\times10^7$~\msun\/ in
molecular and atomic gas (\citealt{israel01}), \citet{harris01} have
identified almost 400 young star clusters from HST WFPC2 imaging. The
ages that they estimate for a subsample of 45 clusters vary between 2
and 47 Myr, with a distribution supporting earlier findings of an age
gradient within the star forming arc, and suggestive of a northward
propagation of the star formation activity (\citealt{gallais91},
\citealt{puxley97}).  The arc of hot spots is found to lie between two
circumnuclear rings, identified from $J$ and $K$ imaging by
\citet{elmegreen98} as the inner Lindblad resonances.
Additional components of this complex region include the optical
nucleus (the brightest source in the near IR), a possible
highly-extinct second nucleus (\citealt{thatte00}), a bar connecting
the hot spot arc to the nucleus along the direction of the main
stellar bar, and additional sites of star formation to the W of the
nucleus, better observed at 10 $\mu$m and at radio wavelengths
(\citealt{telesco88}) because of their high extinction.

In the nuclear region of NGC~3351 ($D$\,=\,10.0 Mpc,
\citealt{freedman01}) several regularly-spaced \hii\/ regions define
an elongated ring $13\times6$ arcsec ($630\times290$ pc) in size
(\citealt{alloin82}). Eight main hot spots can be identified in the
H$\alpha$ images of \citet{planesas97}, each with a luminosity
L($H\alpha$)\,$>$\,$10^{38}$ erg\,s$^{-1}$.  A comparison of the ring
morphology in $H\alpha$, near-infrared (\citealt{elmegreen97},
\citealt{perez00}) and UV (\citealt{colina97}) reveals slight
differences. Moreover, the peaks of the CO emission
(\citealt{kenney92}) show an offset from the $H\alpha$ and $K$-band
peaks. Various explanations exist for this effect, including patchy
extinction from dust, stellar winds wiping out gas and dust, and aging
of the stellar populations (\citealt{conselice00},
\citealt{kotilainen00}, \citealt{maoz01}).

Previous attempts to characterize the stellar populations of
circumnuclear hot spots have suggested the presence of some
peculiarities when compared with disk \hii\/
regions. \citet{kennicutt89} noted that the stronger stellar continua
relative to the emission-line luminosity observed in hot spots could
be a signature of an unusual IMF or, alternatively, could imply a
continuous regime of star formation, rather than an instantaneous
burst activity, as is commonly inferred for disk EHRs.  Continuous
star formation in individual hot spots seems to be ruled out by more
recent studies. \citet{kotilainen00} favor instantaneous burst models
with ages of $\sim$6 Myr to reproduce their Br$\gamma$ equivalent
width measurements (see also \citealt{ryder01}). For the galaxies
examined in this work, hot spot ages of 10 Myr or younger are derived
on the basis of near-infrared colors
(\citealt{elmegreen97,elmegreen98}), Si\,{\sc iv}/C\,{\sc iv}
equivalent width ratios (\citealt{colina97}), and near-infrared CO line
equivalent widths (\citealt{thatte00}). However, unusual stellar mass
spectra have not been firmly ruled out.

In this paper we investigate the stellar content of hot spots and
other metal-rich extragalactic \hii\/ regions. Our work is organized
as follows: we describe the spectroscopic observations and the data
reduction in Section~2, and discuss some of the \hii\/ region global
properties (reddening, density and metallicity) in Section~3. The main
nebular diagnostics are presented in Section~4, and the parameters
derived for the hot spots, in particular their ages, are discussed in
Section~5. In Section~6 we discuss the W-R features identified in five
\hii\/ regions in M83. Finally, a comparison with evolutionary and
photoionization models is carried out in Section~7. Section~8
summarizes the main results of this paper.

\section{Observations and data reduction}

The mappings of the nucleus of M83 by
\citet{elmegreen98} and of NGC~3351 by \citet{colina97} and
\citet{planesas97} were used to define the nuclear targets in the two
galaxies. Inner disk \hii\/ regions were selected from \citet{devau83}
and \citet{hodge74}, respectively. For the third galaxy in our sample,
NGC~6384, \hii\/ region positions were taken from \citet{feinstein97}.
Final target positions and slit orientations were decided on the basis
of short-exposure narrow-band $H\alpha$ acquisition images taken
shortly before the spectroscopic frames, by centering the slit on top
of the emission peaks.

Long-slit spectra of the target \hii\/ regions were obtained with the
EFOSC2 spectrograph mounted at the ESO 3.6m telescope on La Silla on
May 15-16, 1999. The seeing varied between 1.5 and 2.0 arcsec during
the first night, and remained constant around 1.0 arcsec on the
second. The slit used was 2 arcsec wide, also in periods of
unfavorable seeing conditions, as a tradeoff between light-losses and
spectral resolution, and also in order to include most of the emission
from the spatially extended \hii\/ regions.  The spectra were obtained
with three separate exposures through different grisms (no. 7, 9 and
10, all 600 gr/mm) to ensure a continuous coverage of the
3500-8000~\AA\/ wavelength range. The spectral resolution achieved is
approximately 10~\AA, as measured by the full width at half maximum of isolated
emission lines in the reduced spectra. Exposure times varied between
600\,s and 1800\,s, with the longest exposures obtained in combination
with grism no. 9, which covered the spectral range between 4700 and
6770~\AA. Longer exposures times (up to one hour) were chosen for the
three faint \hii\/ regions observed in NGC~6384. Four to five standard
stars from the list of \citet{hamuy92} were observed each night for
flux calibration.

\begin{figure}
\plotone{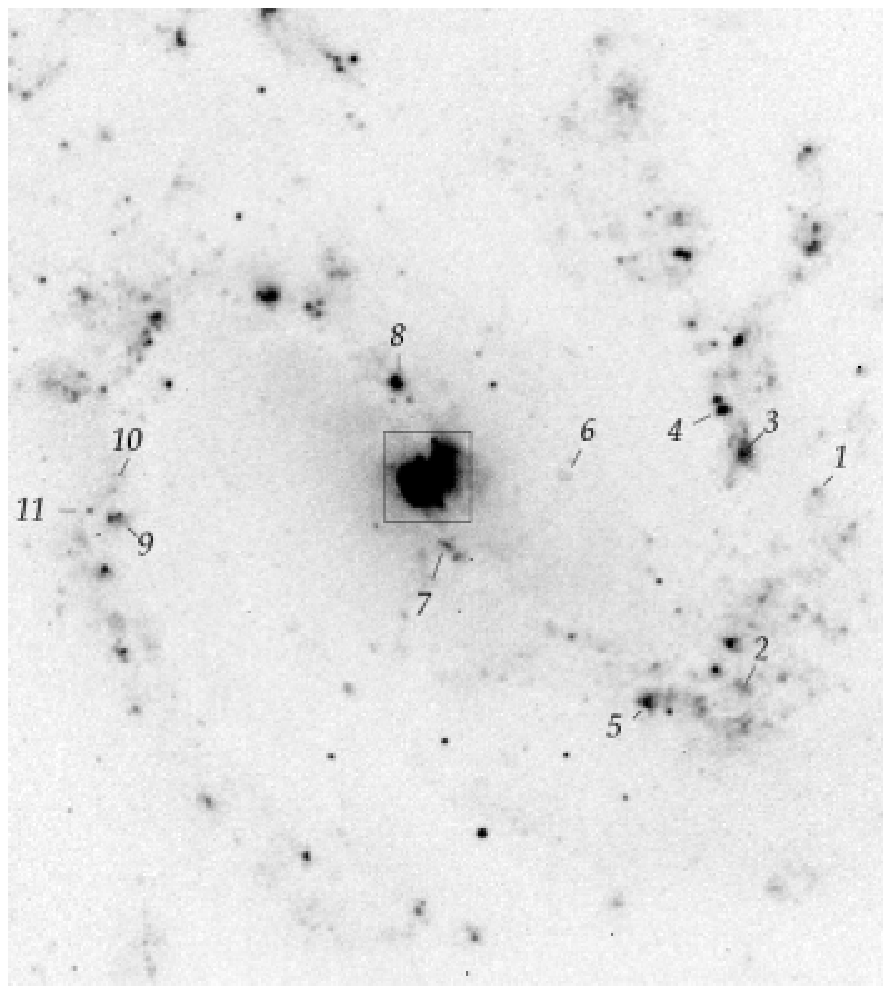}
\caption{$H\alpha$ (line\,+\,continuum) image of M83 identifying the
\hii\/ regions observed. North is at the top, East to the left, the field
of view is $5\arcmin\times5\arcmin$. The central area, indicated by
the square, is shown in Fig.~\ref{2nuc}.\label{m83.fig}}
\end{figure}

\begin{figure}
\plotone{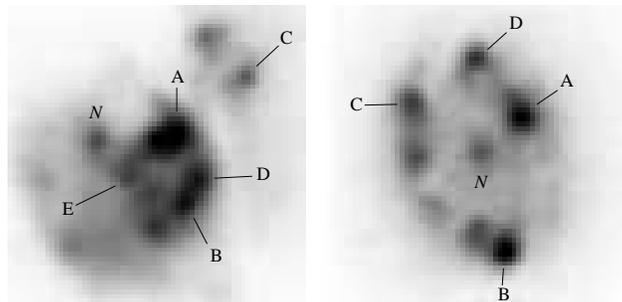}
\caption{$H\alpha$ (line\,+\,continuum) images of the nuclear regions
($20\arcsec\times20\arcsec$) identifying the hot spots observed.
North is at the top, East to the left. {\em a:} M83. {\em b:} NGC~3351.
\label{2nuc}}
\end{figure}

\begin{figure}
\plotone{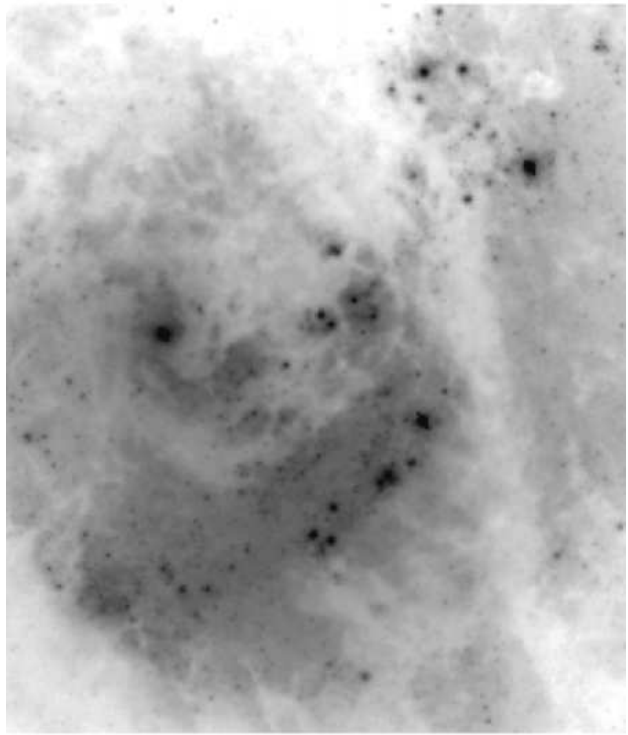}
\caption{Archival HST/STIS visible image of the nuclear region of
M83. Orientation and field of view as in Fig.~\ref{2nuc}a.\label{m83stis.fig}}
\end{figure}

During the observations the spectrograph was rotated to include at
least two \hii\/ regions within the slit simultaneously, while at the
same time keeping the position angle very close to the parallactic
angle.  The final sample includes 16 \hii\/ regions in M83 (of which
five are hot spots), six in NGC~3351 (four hot spots) and three in
NGC~6384. The position of the M83 regions is shown on top of $H\alpha$
aquisition images in Fig.~\ref{m83.fig} for the disk \hii\/ regions
and in Fig.~\ref{2nuc}a for the nuclear hot spots.  An HST archival
(program 7577, PI: Heap) STIS visible image of the same field is shown
in Fig.~\ref{m83stis.fig}, to illustrate the distribution of the star
clusters present in the region at high spatial resolution.  The
nuclear hot spots studied in NGC~3351 are identified in
Fig.~\ref{2nuc}b. Positions relative to the galaxy center and cross
identifications of all the objects are summarized in Table~1. For
convenience, the \hii\/ regions in the disk are indicated by numerals,
and hot spot by letters.

\begin{figure}
\plotone{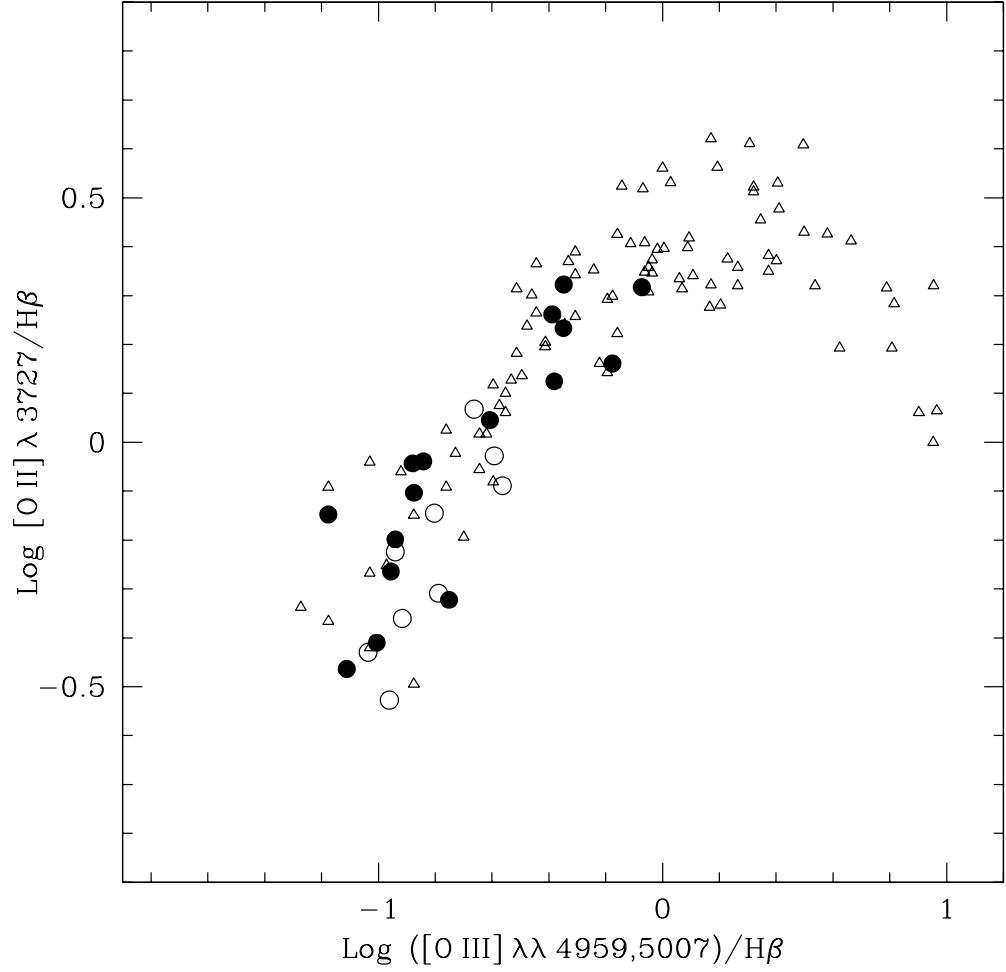}
\caption{The [O\,II] vs.~[O\,III] diagram, illustrating the low excitation
of most of the \hii\/ regions in our sample.  {\em Filled circles:}
disk \hii\/ regions; {\em open circles:} hot spots; {\em open
triangles:} EHR sample from BKG.
\label{oiioiii.fig}}
\end{figure}

The reduction of the long-slit spectra was carried out using standard
IRAF\footnote{IRAF is distributed by the National Optical Astronomical
Observatories, which are operated by AURA, Inc., under contract to the
NSF} procedures, and included bias removal, flat fielding correction,
and wavelength and flux calibration. For the extraction of the
calibrated spectra a suitable region for sky subtraction was
identified, and care was taken to add the flux from the same number of
pixels along the spatial direction in all of the three sets of spectra
taken with different grisms for each given object.  The relative
calibration of spectra of the same object taken with different grisms
was made possible by the emission lines in common, i.e.~$H\beta$
(between grisms no.\,7 and no.\,9) and $H\alpha$\,+\,[N\,II] (between
grisms no.\,9 and no.\,10).  For the emission line measurement we
integrated the flux under the line profile, after identifying the
continuum on both sides of the line. For the
$H\alpha$\,+\,[N\,II]\ww6548,\,6583 and the [S\,II]\ww6716,\,6731 blends
we used the deblending option available with the IRAF {\em splot}
program.  The low excitation of the \hii\/ regions observed,
reflecting their high metal content, is shown by the faintness of the
high-excitation lines relative to $H\beta$, such as [Ar\,III]\w7135 and
[O\,III]\w5007 (the latter has an average $I_{5007}/I_{H\beta} =
0.018$), as illustrated in Fig.~\ref{oiioiii.fig}.

The He\,I recombination lines at 5876~\AA\/ and 6678~\AA\/ were
measured with special care, given their importance as \teff\/
diagnostics in low excitation nebulae \citep{kennicutt00}. In most of
the disk \hii\/ regions the measurement of He\,I\,\w5876 presented no
particular difficulty, but in a few nebulae with stronger continuum,
particularly in most of the hot spots, the emission line is
contaminated by stellar absorption features, most notably the Na\,I D
line at 5890~\AA\/ (which can also have an important interstellar
component). A comparison with stellar spectral atlases shows that the
main responsible for these features are late-type giants, and their
presence can make the measurement of He\,I\,\w5876 in emission very
uncertain. In such cases we preferred to measure only the \w6678 line,
when present, or to estimate an upper limit for it, on the basis of
the random noise fluctuations in the continuum adjacent to the
line. An equivalent width of 0.5~\AA\/ was assumed for the stellar
absorption component of the He\,I lines.

\begin{figure} 
\plotone{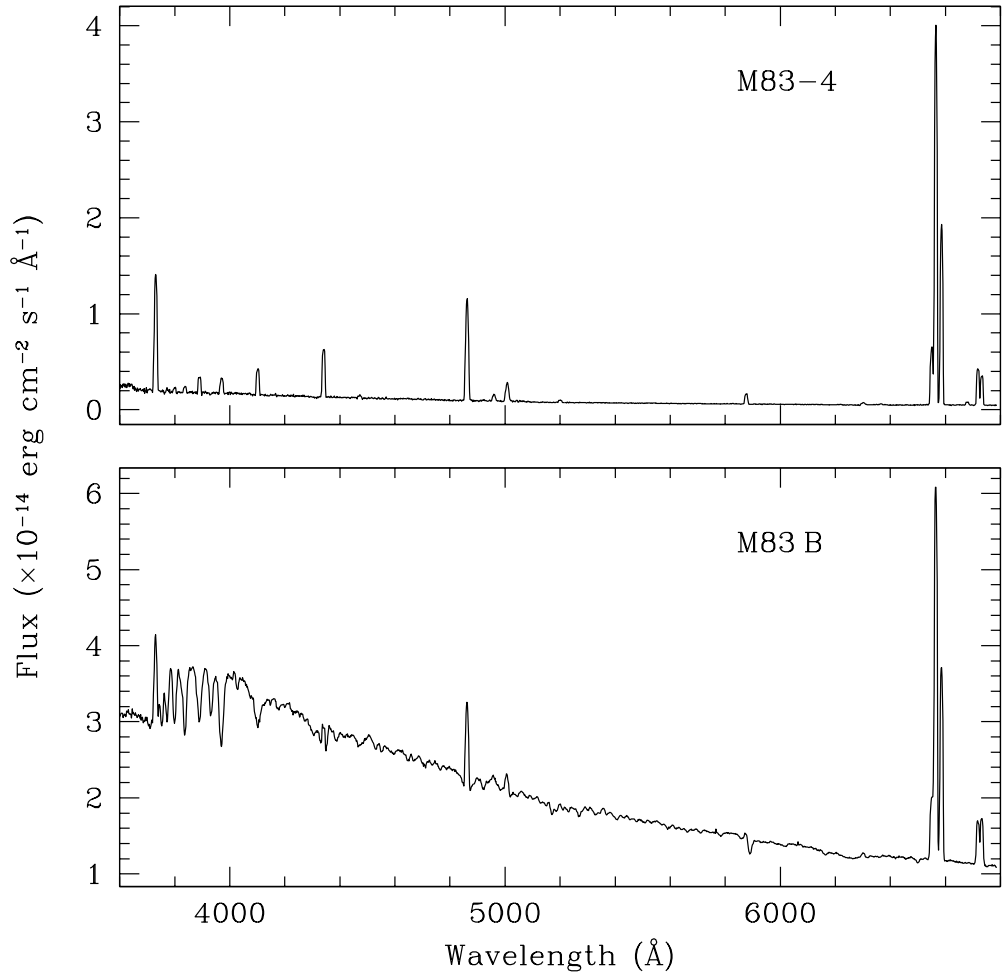} 
\caption{Examples of
\hii~region spectra, corrected for extinction, representative of the
disk and nuclear regions in the current sample. (Top) Disk region 4 in
M83, showing an almost pure emission line spectrum on top of a weak
continuum. (Bottom) Hot spot B in M83, with nebular lines on top of a
strong continuum, and prominent absorption features characteristic of
the underlying stellar population.
\label{HIIexamples.fig}}
\end{figure}

For the computation of the extinction by dust by means of the
Balmer decrement (\citealt{osterbrock}) the contamination of the
emission line spectra by an underlying stellar absorption
component must be taken into account.  The presence of stellar
absorption features is particularly evident in the case of the
hot spots, where, with respect to what is normally found in
EHRs, an additional stellar component is seen, because of the
contribution to the total light coming from the bulge stellar
population. This is illustrated in the lower plot of
Fig.~\ref{HIIexamples.fig}, which displays two representative
spectra from our \hii\/ region sample, one being an M83 disk
\hii\/ region with low continuum, the second a nuclear
hot spot (M83\,B) showing, in addition to the higher-order Balmer and
He\,{\sc i} lines in absorption, originating from young O, B and A
stars, also a series of prominent metallic absorption features from
the older stellar population in the galactic bulge.  The most
conspicuous are the K~Ca\,{\sc ii}\w3933\,\AA, the G band at
4301\,\AA\/ and the Mg\,{\sc i}\,+\,MgH\w5175\,\AA\/ lines. Their
equivalent widths are useful as metallicity and age indicators of
galactic stellar populations (\citealt{bica87},
\citealt{bica88}), and are often used, with the aid of
appropriate nuclear galaxy spectral templates among those published by
\citet{bica88}, to measure the contribution of different age and
metallicity components to the integrated light of galaxies and
clusters. Recent applications of stellar population synthesis using a
base of star clusters can be found in \citet{raimann00} and
\citet{saraiva01}. We have compared our equivalent width values 
(See Table~2) with those reported by \citet{bica88} for different
spectral templates, and found that they are in most cases considerably
lower than those of his class S7, which has the smallest equivalent
widths for the metallic lines, indicating a considerable dilution by
the continuum from the young stars.  The integrated values measured in
the central 10\arcsec $\times$ 20\arcsec\/ of M83 and NGC~3351 (thus
including the circumnuclear ring of star formation) by
\citet{storchi95} are also a factor 1.4 to 3.8 larger than those found
here, as a result of the bulge contribution included in the larger
aperture.  Lacking an appropriate template for the comparison, we
opted for the traditional method of deriving the contribution for the
underlying stellar population (see
\citealt{mccall85}), for both the disk \hii\/ regions, as well as the
hot spots. An iterative procedure was used to determine what values of
the underlying equivalent width of the Balmer lines in absorption
($W_{abs}$), assumed to be equal for all Balmer lines, allowed us to
recover simultaneously the case B~$H\alpha$/$H\beta$ and
$H\gamma$/$H\beta$ emission line ratios (\citealt{hummer87}) for a
given reddening \ebv, assuming $T_e=5000~K$ and the
\citet{seaton79} Galactic extinction law.  The $W_{abs}$ values
thus derived span a wide range, from 0~\AA\/ up to 4.4~\AA, with the
hot spots not showing particularly large values when compared to the
disk EHRs.  The absorption equivalent widths are summarized in
Table~3 (Column~3), together with the reddening-corrected line fluxes
for the measured lines, normalized to $H\beta$\,=\,100 (Columns~4-12),
the reddening \ebv\/ (Column~2) corrected for the Galactic foreground
(based on \citealt{schlegel98}) and the $H\beta$ emission line
equivalent widths (Column~13).  In the case of emission lines which
are theoretically expected to be in well-defined ratios to brighter
lines, i.e.~[O\,III]\w4959 and [N\,II]\w6548, we report only the
measurement of the brightest line of the pair, namely [O\,III]\w5007
and [N\,II]\w6583.  Uncertainties in the tabulated values are
estimated to be 5\% for the bright lines ([O\,II]\w3727,
[O\,III]\w5007, He\,I\,\w5876, [N\,II]\w6584, [S\,II]\ww6717,6731), and
between 10\% and 20\% for the fainter lines ([O\,I]\w6300,
He\,I\,\w6678, [Ar\,III]\w7135).

\section{Analysis of global properties}

\subsection{Reddening}
Looking at Column~2 in Table~3, we notice no significant difference in
\ebv\/ between disk \hii\/ regions (average 0.36) and hot spots
(average 0.41). The mean value
\mbox{\ebv$\;\simeq\;$0.38} is comparable to that measured by \citet{ho97}
in \hii\/ regions located in galactic centers, and which they found to
be not significantly larger than the average reddening in the EHRs
studied by \citet{mccall85}. \citet{oey93} considered EHRs in
early-type spirals, finding that their typical metallicity (mean
$12+\log (O/H)=8.97$), based on the $R_{23}$ calibration of
\citet{dopita86}, is a factor 1.5 larger than for the objects in the 
\citet{mccall85} sample. Nevertheless, the reddening in the two 
samples is comparable, being in fact somewhat smaller in the
early-type galaxies. The relatively low value of \ebv\/ in the nuclear
zone of M83, despite its high metallicity, was already noted by
\citet{storchi94}, indicating that other factors beside metal
abundance determine the amount of extinction, such as the galaxy
inclination to the line of sight and the geometrical distribution of
dust within the star forming regions.

In the \hii\/ regions located in the proximity of the M83 nucleus
heavy and patchy extinction by dust ($A_V>13$ mag) is inferred from
mid-IR observations \citep{turner87, rouan96}. In the $K$ band
\citet{thatte00} derived a wide range of extinction values, from 0.5
mag up to 9.2 mag. The fact that the optically derived extinctions are
much lower (generally $<$\,2 mag) is an indication of the various
depths probed at different wavelengths, the mid-IR reaching far more
into the star forming regions then the near-IR and the optical, and
that the optically measured regions are very heavily weighted to
regions with the lowest extinction. A similar effect has been found in
other nearby starbursts, as in, for example, He2-10
(\citealt{beck97}).

While there seems to be no direct evidence for a connection between
metallicity and amount of extinction as derived from the Balmer
decrement, it is legitimate, however, to ask the question whether dust
can affect the emission-line spectra of metal-rich \hii\/ regions in
other ways, especially if the gas-to-dust ratio is increasing with
metallicity.  The work by \citet{shields95} in fact showed that the
depletion of heavy elements onto dust grains can have measureable
effects on the emission line spectra of metal-rich \hii\/ regions. The
removal of gas coolants and the selective absorption of continuum
photons at wavelengths longer than the Lyman limit combine to increase
the equilibrium temperature of nebulae at high metallicity, with the
result that low-ionization emission lines are enhanced with respect to
the dust-free case. Since the impact of this effect is expected to be
stronger in the high metallicity regime, this would provide a natural
explanation for the enhanced [N\,II] and [S\,II] line intensities
measured in \hii\/ regions in the nuclei of normal galaxies
(\citealt{kennicutt89}; \citealt{ho97}). As we will see in
Section~\ref{diagnostics}, a weak effect might be present in the
low-ionization lines measured in the hot spots in M83 and NGC~3351,
which could be attributed to depletion by dust.

\subsection{Density}

\begin{figure}
\plotone{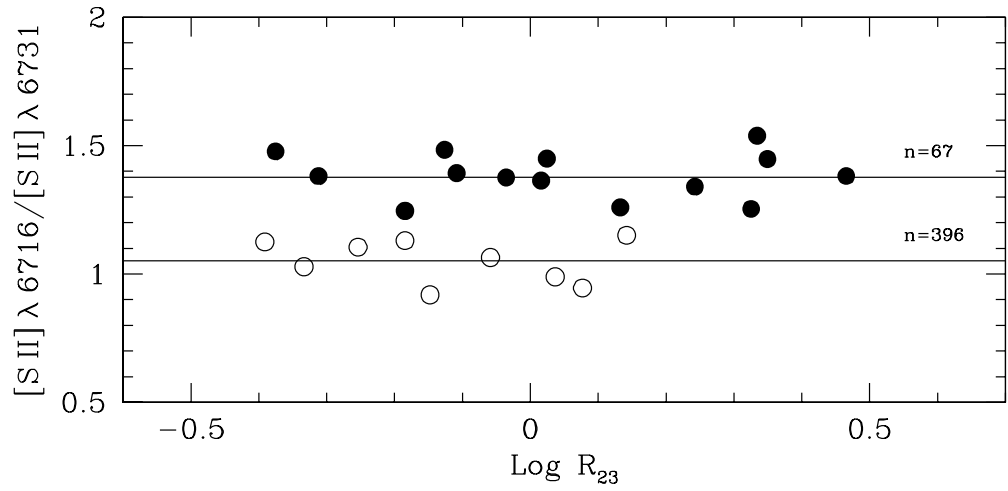}
\caption{The density-sensitive ratio [S\,II]\w6716/[S\,II]\w6731 as a
function of the empirical abundance indicator
$R_{23}$=([O\,III]\ww4959,5007+[O\,II]\w3727)/$H\beta$. The straight
lines are linear fits to the two sequences of disk and hot spot \hii\/
regions. The indicated values for the electron density (in cm$^{-3}$)
were derived from the formulae of McCall (1984).\label{sii}}
\end{figure}

The analysis of the density-sensitive ratio
[S\,II]\w6716/[S\,II]\w6731 shows that disk \hii\/ regions and hot
spots are well separated into two distinct sequences, as seen in
Fig.~\ref{sii}, which plots the [S\,II] doublet ratio as a
function of $R_{23}$.  The electron density in hot spots, derived
from the observed ratios using the analytical formulae of
\citet{mccall84}, is on average approximately 400 cm$^{-3}$,
about six times larger than in disk \hii\/ regions. The average value
for the latter is in good agreement with that found in the larger EHR
sample of BKG. Such a difference is in general expected to have only a
secondary effect on nebular emission properties, which explains why
the diagnostic diagrams such as those in Fig.~\ref{baldwin} (next
Section) show no clear separation of disk and circumnuclear \hii\/
regions. It has been shown by \citet{oey93}, however, that at high
metallicity density can have an important effect on the collisional
de-excitation of the $O^{++}$ fine-structure lines, which leads to an
increased uncertainty in the $R_{23}$ abundance scale. The
metallicities derived in the following section must therefore be taken
with some care, although the relative values should still be reliable.

\subsection{Metallicity}
The question of how metal-rich the objects in our sample are does not
have a simple and unique answer. These EHRs belong to the lowest end
of the extragalactic \hii\/ region excitation sequence (see
Fig.~\ref{oiioiii.fig}), and as a result of the increased metal
abundance most of the cooling of the nebular gas occurs via fine
structure infrared lines, such as the [O\,III]\ww52,88 $\mu$m
lines. The consequent unobservability of the $T_e$-sensitive auroral
lines in the optical forces us to use semi-empirical or theoretical
approaches to estimate the chemical composition. Several
semi-empirical abundance techniques can be found in the literature,
employing different ratios of relatively bright nebular lines. For the
widely used $R_{23}$, originally proposed by
\citet{pagel79}, different authors have provided calibrations which
might differ systematically at the high-metallicity end ($\log
R_{23}$\,$<$\,0) by up to 0.2-0.3 dex (\citealt{ep84};
\citealt{mccall85}; \citealt{dopita86}, see the comparison in \citealt{mcgaugh91}). 
Additional empirical methods have been proposed, including
[O\,III]\w5007/[N\,II]\w6583 (\citealt{alloin79},
\citealt{dutil99}), [N\,II]/$H\alpha$ (\citealt{vanzee98}), and
[N\,II]/[O\,II] (\citealt{vanzee98}, see also
\citealt{dopita00}). In general, however, at high metallicity (solar and above)
the calibration of these methods depends on a scant number of
observations and on photoionization modeling (for example, the M101
\hii\/ region Searle 5 in the case of the \citealt{ep84} work), and their
reliability is often questioned. Typical uncertainties of 0.1-0.2 dex
are normally quoted for abundances derived with these empirical
methods at high metallicity, but the systematic differences mentioned
above imply an uncertain absolute abundance scale, dependent on the
choice of the semi-empirical calibration. Obviously, any comparison
with metal-rich \hii\/ regions in the literature must take this choice
into account. Recent work on metal-rich EHRs with measureable auroral
lines by \citet{castellanos01} has shown indications that empirically
determined abundances might be severely overestimated, by up to a
factor of three (see also
\citealt{pyliugin01b}). The upper branch of the $R_{23}$ vs.~$O/H$
relation (their Fig.~9), based on their new data and on the theoretical
modeling of M51 \hii\/ regions by \citet{diaz91}, has either a flatter
slope or a systematically lower zero-point than the commonly
adopted semi-empirical calibrations, such as those of
\citet{mccall85} or \citet{ep84}.

In this paper we adopt the analytical expression reported by
\citet{kobulnicky99}, which is based on the photoionization models
calculated by \citet{mcgaugh91}, and in which both
[O\,III]\ww4959,5007/[O\,II]\w3727 and $R_{23}$ are used to
simultaneosuly contrain $O/H$ and the ionization parameter $U$.  This
calibration is in good agreement with the experimental data and the
photoionization models shown by
\citet{castellanos01} at high abundance, and has an estimated uncertainty
of 0.1 dex.  The McGaugh models show a relative insensitivity to the
ionization parameter $U$ and to the stellar effective temperature \teff\/
in the metal-rich regime. Empirically, at high metallicity the spread
in $U$ is found to be rather small, with high-abundance EHRs having,
on average, $\log U\simeq -3$ (\citealt{vanzee98}), or a
somewhat smaller value, based on the [S\,II]/[S~III] line ratio ($\log
U\simeq -3.4$, from the BKG data), or the
[O\,II]/[O\,III] and [S\,II]/$H\beta$ ratios of the current sample,
based on the calibrations given by \citet{diaz00} and
\citet{gonzalez99c}, respectively ($\log U\simeq -3.5$).

There are some additional caveats to consider in the use of the strong
line method at high abundance, as summarized by
\citet[cf.~\citealt{stasinska01b}]{stasinska99}. The
\citet{mcgaugh91} models assume unevolving ionizing star clusters,
whereas the effects of evolution on the oxygen line strengths
predicted by synthetic models are rather profound at high
metallicity. This is due to the extreme sensitivity of the electron
temperature and the [O\,III]\w5007 line intensity to changes in
nebular geometry, density and mean effective temperature in this
abundance regime, making the strong line method very uncertain at
supersolar metallicities. Abundances based on the auroral lines will
be even more significantly affected.  That the thermal equilibrium of
metal-rich nebulae is affected by the electron density was pointed out
in the study of high-abundance \hii\/ regions of \citet{oey93}. On the
other hand the boosting of the [O\,III]\w5007 line due to the
appearance of W-R stars could be overestimated in the models, since
the observations (see Section 7) show that \teff\/ is hardly affected
by the presence of W-R stars.

We have verified that all of our \hii\/ regions belong to the upper
branch of the $R_{23}$-abundance relationship, by using the
[N\,II]/[O\,II] line ratio, which monotonically increases with
abundance (\citealt{mcgaugh94}).  The oxygen abundances $12+\log
(O/H)$ determined in this way are listed in Column 14 of Table~3.  A
comparison with existing semi-empirical calibrations show
significative systematic differences. For example, the mean oxygen
abundance of our sample using the McGaugh theoretical calibration is
$12+\log (O/H)$\,=\,9.05.  The $R_{23}$ calibration by
\citet{ep84} gives a mean of 9.31, but the revised calibration at high
abundance by \citet{edmunds89}, adopted in the abundance gradient work
of \citet{vilacostas92}, lowers it to 9.14, more in agreement with the
McGaugh calibration. The two methods agree at the lower abundance end
(around solar), while at the higher end they differ by up to $\sim$0.2
dex.  As a comparison, both the abundances obtained for our sample \hii\/
regions using the two calibrations are given in the last two
columns of Table~3 (MG for \citealt{mcgaugh91}, and E for
\citealt{edmunds89}). Considering additional bright line methods, 
the [O\,III]/[N\,II] line ratio calibrated by \citet{dutil99} gives a
mean abundance $12+\log (O/H)$\,=\,9.20, and the $P$-method of
\citet{pyliugin01} provides an average of 8.76. 

\begin{figure}
\plotone{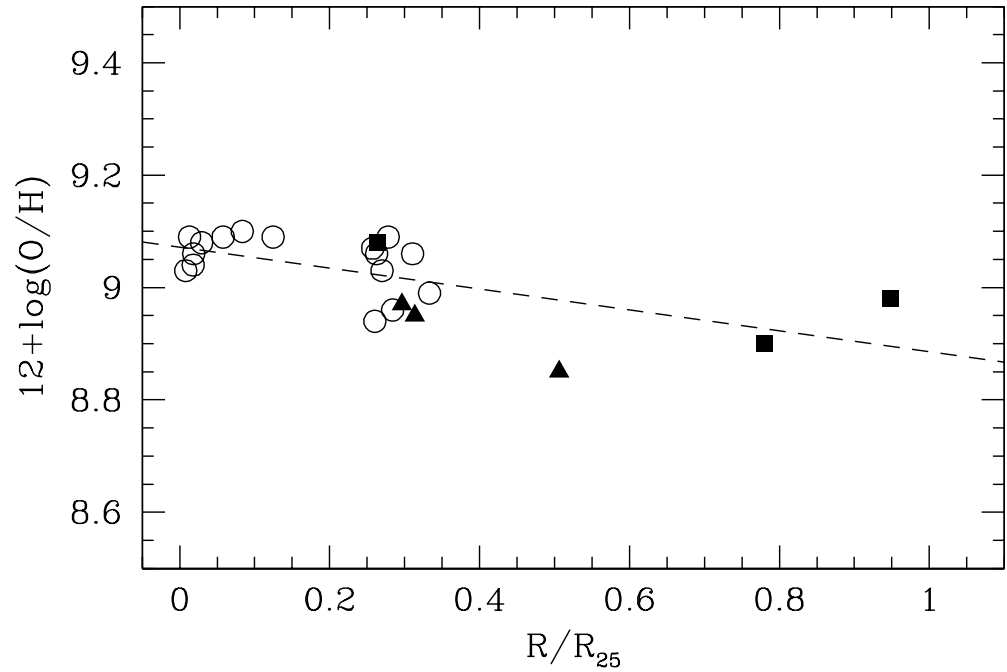}
\caption{The M83 \hii\/ region oxygen abundance as a function of
$R/R_{25}$ (using $R_{25}=6\farcm74$) from our data (open
circles), and from observations of additional \hii\/ regions by
\citet[squares]{dufour80} and \citet[triangles]{webster83},
adopting the \citet{kobulnicky99} strong line method
calibration. The dashed line is a least square fit to all the data
points, with intersect 9.07 and slope -0.186.
\label{M83gradient}}
\end{figure}

As shown in previous work in M83, the \hii\/ regions trace a shallow
radial abundance gradient, and Table~3 shows on average slightly
higher abundances in the inner \hii\/ regions and the hot spots than
in regions further out. The radial coverage is however too small for a
determination of the overall radial abundance gradient in this galaxy.
With the adopted calibration the abundance in the nucleus is 
$Z$\,$\simeq$\,1.5\,\zsun.  An early investigation of six
\hii\/ regions in M83 by \citet{dufour80} found much higher values, up
to 4-5\,$Z_\odot$, from a differential analysis based on theoretical
models of the [O\,II] and [O\,III] lines. A comparison of the
observational data with more recent grids of
\hii\/ region models (e.g.~\/\citealt{dopita00}) suggests considerably lower
values, on the order of 2\,$Z_\odot$, consistent with our result.  The
central value quoted by
\citet{vilacostas92}, based on a fit to the optical data of
\citet{webster83} and
\citet{dufour80}, is $12+\log (O/H)$\,=\,9.24 ($\sim$2.2\,\zsun). A similar
result is obtained from the work of \citet{zaritsky94}.  Considering
the differences in empirical calibrations previously discussed, this
is in rough agreement with our result. The radial abundance gradient
traced by the \hii\/ regions observed by us and from previous work is
illustrated in Fig.~\ref{M83gradient}. Here the oxygen abundances from
the data published by \citet[squares]{dufour80} and
\citet[triangles]{webster83} have been calculated using
the same \citet{kobulnicky99} empirical calibration adopted for our
data.

In NGC~3351 the uniformity in abundance for the observed \hii\/
regions is consistent with previous studies
(e.g.~\/\citealt{zaritsky94}, \citealt{dutil99}), which measured no
significative abundance gradient in the inner parts of this
galaxy. The central value from our measurements and the adopted
abundance calibration is $Z\simeq$\,1.6\,\zsun\/ (2.6\,\zsun\/ with the
\citealt{edmunds89} calibration).  Finally, the three objects observed
in NGC~6384 have a roughly solar abundance, in agreement with the work
of \citet{oey93}.

\section{Diagnostic diagrams}\label{diagnostics}

\begin{figure}
\plotone{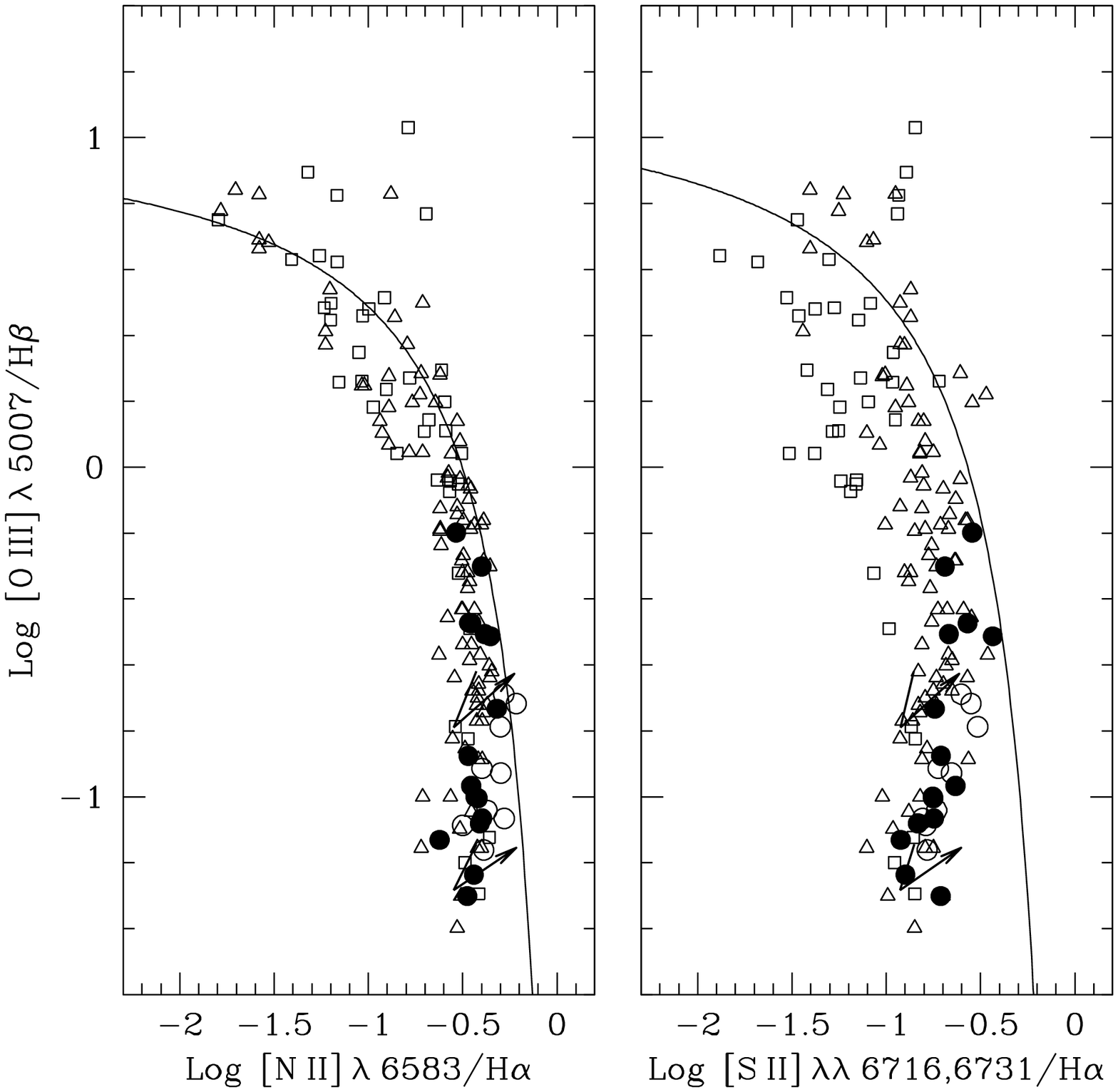}
\caption{Nebular excitation sequence shown as [N\,II]/$H\alpha$ and
[S\,II]/$H\alpha$ vs [O\,III]/$H\beta$. {\em Filled circles:} disk
\hii\/ regions; {\em open circles:} hot spots; {\em open squares:}
galactic \hii\/ region sample from \citet{kennicutt00}; {\em open
triangles:} EHR sample from BKG. The theoretical upper boundaries for
a zero-age instantaneous burst of
\citet{dopita00} are shown for comparison. The four Galactic objects
well above this boundary are \hii\/ regions ionized by W-R stars, the
extragalactic one is M\"{u}nch~1 in M81. The effects of changing the
density from 70 to 400 cm$^{-3}$ (shown by the steep line segment) and
of dust (arrow) are indicated at the position calculated with solar
composition Cloudy photoionization models at $n_e$=70 cm$^{-3}$, $\log
U=-3$, and CoStar stellar atmosphere models (\teff\,=\,36,000~K and
\teff\,=\,37,000~K).
\label{baldwin}}
\end{figure}

We present here some general properties of our \hii\/ region
sample, as derived from some standard diagrams which are
recognized as helpful diagnostics of the nebular physical
conditions. These are often used to help discriminate between
nebulae photoionized by massive stars and those involving
nonstellar mechanisms (i.e.~AGNs, \citealt{veilleux87}). Recent
applications of this technique can be found, among others, in
\citet{vanzee98}, \citet{martin99}, BKG, \citet{kennicutt00},
\citet{barth00}, and \citet{contini01}. The excitation sequence
is shown in terms of [N\,II]/$H\alpha$ vs.~[O\,III] and
[S\,II]/$H\alpha$ vs.~[O\,III] in Fig.~\ref{baldwin}. In this and in
all following plots we have used filled circles for disk
\hii\/ regions and open circles for circumnuclear hot spots. The
squares represent the data points taken from the galactic \hii\/
regions studied by \citet{kennicutt00}, while the triangles represent
the EHR sample of BKG, shown here as comparison samples.  The two
sequences in Fig.~\ref{baldwin} show how our new extragalactic sample
well matches the low-excitation, low-\teff\/ end of the galactic
\hii\/ region sequences and of the known sequences of EHRs. This
indicates that young and massive stars are the primary photoionization
source for the whole sample. The plots suggest that the hot spots
might have a somewhat enhanced [N\,II] and [S\,II] emission, as
predicted by the \citet{shields95} models, but the effect, if real, is
very small ($\sim$0.1 dex).  This is in contrast with the results
obtained by \citet{ho97} in nuclear \hii\/ regions, which show clear
signs of enhanced emission in the low-ionization lines (see also
\citealt{kennicutt89}).   In the specific case of the hot spots,
we reach the conclusion that they are, generally speaking, virtually
indistinguishable from disk EHRs in terms of their emission line
spectra, and, although located in regions of somewhat extreme physical
conditions like the nuclear zone of spiral galaxies, they do not share
some of the anomalous properties of \hii\/ nuclei nor do they show
strong effects due to dust at high metallicity.

To verify this interpretation we have calculated a small set of simple
photoionization models with Cloudy (Version 94.0,
\citealt{ferland97}). CoStar atmosphere models at solar
metallicity and interpolated for stellar temperatures
\teff\,=\,36,000~K and \teff\,=\,37,000~K (\citealt{schaerer97}) provided
the ionizing fluxes. To gauge the effects of density and depletion by
dust on the predicted emission line fluxes, we calculated models at
two different densities, $n_e$\,=\,70 and $n_e$\,=\,400 cm$^{-3}$, to match
the values derived from the sulphur doublet for disk and hot spot
\hii\/ regions, respectively.  The average depletion factors of the
ISM and the silicate-graphite grain properties were taken to be the
defaults allowed by Cloudy (see
\citealt{ferland97} for further details). The ionization
parameter was set to $\log U=-3$ for a spherical geometry. The effects
of dust on the $n_e$\,=\,70 cm$^{-3}$ models at the two stellar
temperatures are shown by the two arrows, while the short, steeper
lines show the effect of changing the electron density from 70 to 400
cm$^{-3}$. While the displacement in the plots due to density changes
occurs mostly along the excitation sequence, dust works in the sense
of increasing the [N\,II] and [S\,II] emission, as shown by
\citet[cf. also the model predictions in the work of
\citealt{barth00}]{shields95}. The observed line intensity ratios in the hot
spots seem therefore compatible with only a mild effect by dust
depletion.

\begin{figure}
\plotone{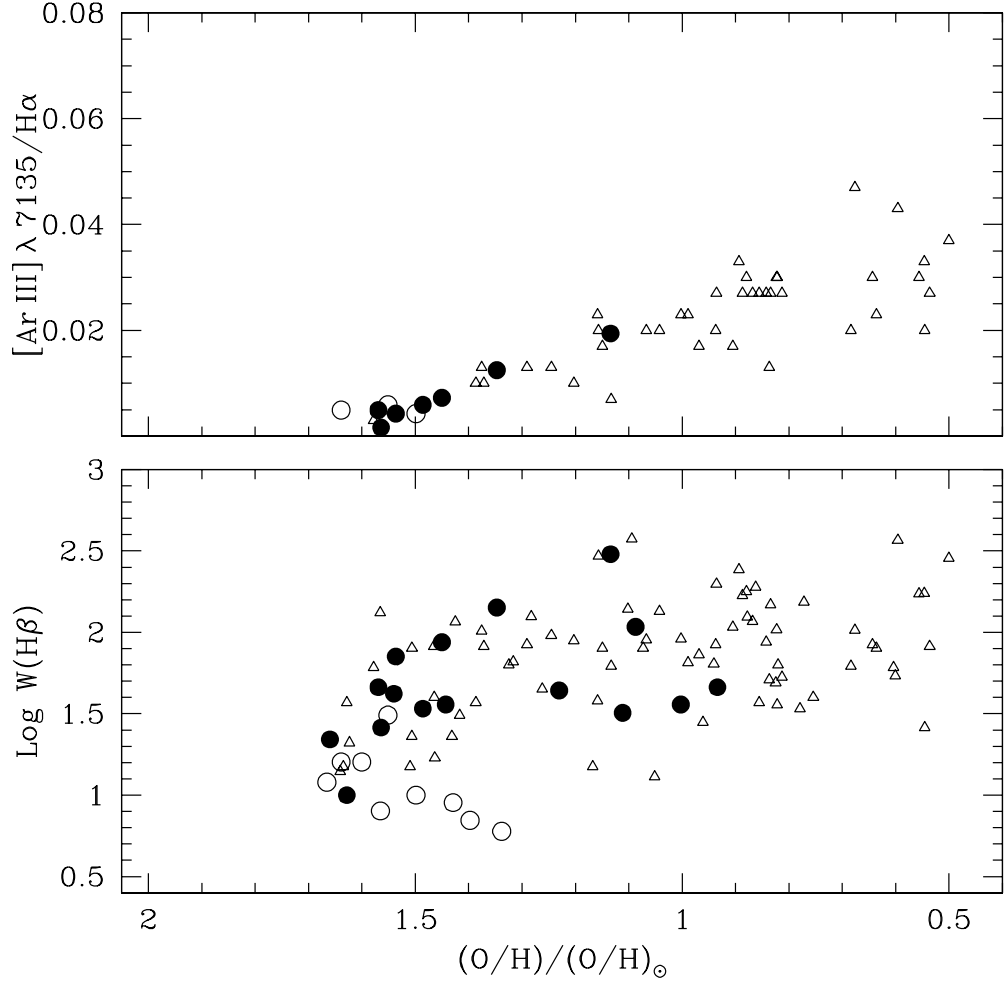}
\caption{(a) The \teff\/-sensitive ratio [Ar\,III]\w7135/$H\alpha$
as a function of the metal abundance relative to solar, estimated with
the strong line method and the analytical calibration given in
\citet{kobulnicky99}. Circles represent the current sample of \hii\/ regions,
open triangles the extragalactic sample of BKG (only objects in the
$R_{23}$ upper branch are plotted).  (b) The equivalent width of the
$H\beta$ emission line as a function of abundance. Symbols as
above.\label{ar}}
\end{figure}

Fig.~\ref{ar} shows the observed trend of two diagnostics of the
ionizing stellar population, [Ar\,III]\w7135/$H\alpha$ and the
equivalent width of $H\beta$, \ew, as a function of $O/H$. The
additional data points (open triangles) are taken from the EHR sample
measured by BKG.  Less than half of the objects in our new sample had
a measurable argon emission line, but the overall trend shown in
Fig.~\ref{ar}(a) is indicative of a clear excitation gradient with
abundance.  A weak trend is also seen in the lower plot, which shows
\ew, corrected for the underlying stellar absorption, vs.~abundance. 
The larger scatter shows the well-known sensitivity of
\ew\/ to the age of the stellar population responsible for the
ionization of the gas, in addition to the ionizing radiation field (in
an evolving star cluster these effects are linked to one another). The
data points, however, suggest a rather well-defined upper boundary in
the plot, which likely corresponds to the location of the youngest
objects, since \ew\/ is predicted to decrease with time (and with
increasing metallicity) by current evolutionary models
(\citealt{cervino94}, \citealt[hereafter SV98]{schaerer98}).

\begin{figure}
\plotone{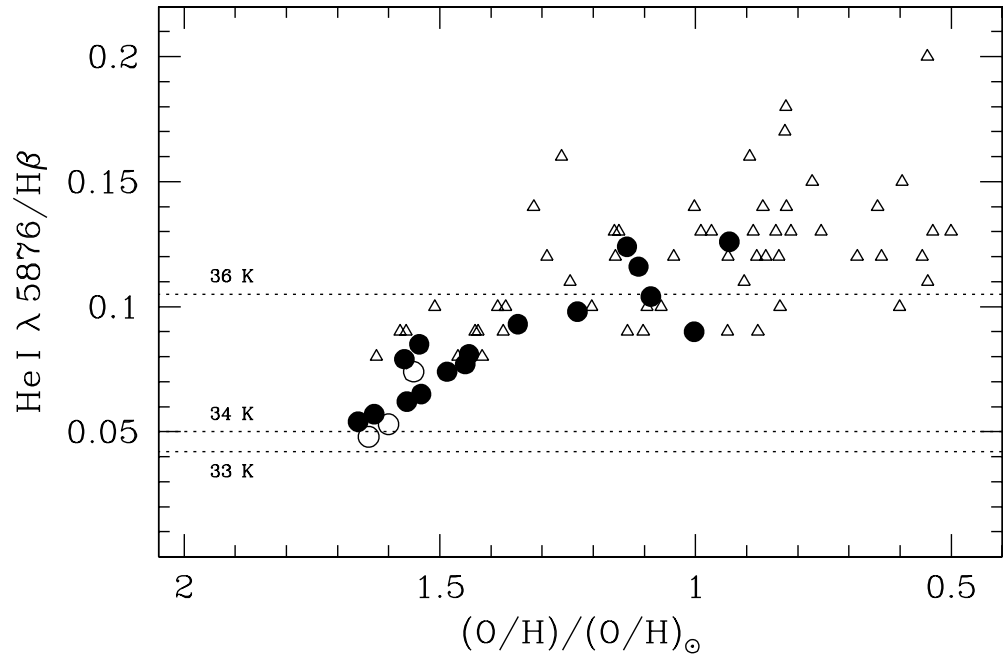}
\caption{He\,I\,\w5876 line intensity relative to $H\beta$ for the \hii\/
regions in our sample (circles) and in the sample of BKG (open
triangles, upper branch of $R_{23}$), plotted as a function of relative solar abundance.  The
horizontal lines, marking \teff\,=\,33, 34 and 36$\times 10^3$~K, have
been drawn based on the empirical calibration of emission line
intensities of
\citet{kennicutt00}.  
\label{he}}
\end{figure}

The presence of a \teff\/-metallicity gradient is perhaps best
estimated (for \teff~$<$~36-38,000~K) by plotting the intensity of
neutral helium lines relative to $H\beta$ as a function of $O/H$, as
we have done in Fig.~\ref{he}, where the data for the He\,I\,\w5876
lines are shown.  The empirical He\,I~line calibration as a function
of \teff\/ from
\citet{kennicutt00} provided the data for the horizontal lines, which
approximately correspond to
\teff\,=\,33,000, 34,000 and 36,000~K. At \teff~$>$~38,000~K the
optical He\,I~lines do not provide information on \teff, due to their
saturation (all helium becoming ionized,
\citealt{doherty95}). According to the diagram in Fig.~\ref{he}
\teff\/ decreases to $\sim$34,000~K at about 1.5 times the solar
abundance, but the absolute abundance scale is subject to the
uncertainties mentioned previously. A similar trend was shown by BKG,
although with a smaller number of extragalactic
\hii\/ regions in the high-metallicity regime. Their He\,I\,\w5876
measurements are represented by the open triangles in Fig.~\ref{he},
where only data belonging to the upper branch (high metallicity) of
the $R_{23}$-abundance relation are shown. In Section 7.4 we will
comment on the implications of this result for the properties of the
ionizing cluster stellar population.

\section{Properties of the nuclear hot spots}

We have established in the previous section that the hot spots can be
regarded as being mostly `normal' EHRs when we consider their
emission-line properties. We analyze here their stellar content in
some further detail.

As mentioned in Section~1, the hot spots in M83 and, to a lesser
extent, those in NGC~3351, have been fairly well studied with
infrared imaging. The complex morphology of the M83 nuclear region in
the near-IR, which is similar to the one in the optical, has been
described by \citet{gallais91} and \citet{elmegreen98}, who have
identified the galaxy nucleus and the arc of star-forming regions 7
arcsec SW of the nucleus as the main emission sources. At longer
wavelengths the picture is however very different, with two bright
sources located in a bar-shaped feature appearing NW of the nucleus at
10~$\mu$m and 6~cm \citep{telesco88}, both having a faint counterpart in
the visible and in the near-IR. Our optical spectra were centered on
$H\alpha$ peaks corresponding to some of the near-IR peaks studied by
\citet{elmegreen98}, and Table~1 provides the identification number
from their paper in the last column.  The mass of the individual
nuclear \hii\/ regions has been estimated in the range
1-4\,$\times10^6~M_\odot$ in M83
\citep{elmegreen98} and 1-10\,$\times10^5~M_\odot$ in NGC~3351
\citep{elmegreen97} from their near-IR colors.  These authors
point out that the regular spacing of the hot spots along
circumnuclear rings is consistent with models of large-scale
gravitational instabilities. \citet{gallais91} and
\citet{rouan96} also suggested that star formation propagating
along the circumnuclear ring of \hii\/ regions in M83 could be a
possible explanation for the observed trend of the $J-H$ and $H-K$
colors along the ring itself. This would imply an age gradient along
the star forming arc. It would be interesting to test this idea with
the current optical spectra, however the determination of \hii\/
region ages from optical spectra can be rather uncertain. We have
considered two different age indicators from the nuclear hot spot
spectra: the equivalent width of the $H\beta$ emission line, and the
profiles of the Balmer series absorption lines.

In Table~3 one hot spot (M83\,A), which corresponds to the position of
region no.~8 in the $K$-band image of \citet{elmegreen98}, stands out
among all others in the current sample by having the largest $H\beta$
equivalent width, \ew\,=\,31\,\AA, as opposed to values beteween 6 and
10~\AA\/ for the remaining four in M83, and 8 to 16~\AA\/ for the four
in NGC~3351.  The small range observed in \ew\/ suggests that all of
the hot spots, except M83\,A, have approximately the same age. The
comparison with the high-metallicity (2\,\zsun), instantaneous-burst
models of SV98 ($M_{up}$\,=\,120\,\msun, Salpeter slope) indicate an
age of 5 Myr for M83\,A and 7 Myr for the remaining objects. A younger
age for M83\,A would be consistent with the possible detection of W-R
stars in it (see Section~6), and with the UV spectra discussed
below. The circumnuclear \hii\/ regions in NGC 3351 appear somewhat
younger then the M83 hot spot regions on average, around 6 Myr. Also
note that in both galaxies only for the younger objects
(\ew\,$\geq$\,16\,\AA) we detect the He\,I\,\w5876 emission
line. According to the results of photoionization models coupled with
evolutionary models (\citealt{gonzalez99c}, \citealt{stasinska01b}) in
the instantaneous burst scenario the forbidden lines, such as
[O\,III]\w5007, and the He\,I recombination lines (He\,I\,\w4471,
\w5876) are expected to fade as the ionizing clusters evolve
(monotonically for the He\,I lines at all metallicities), becoming
unobservable about 6-7 Myr after the onset of the star formation
episode, as a result of the softening of the radiation field after the
highest mass stars have disappeared. In a continuous regime of star
formation, on the contrary, these lines would maintain an equilibrium,
roughly constant ratio to the $H\beta$ emission line, close to the
zero-age value. The detection of He\,I\,\w5876 only for larger \ew\/
values might therefore be interpreted as an argument in favor of the
burst mode of star formation.

\begin{figure}
\plotone{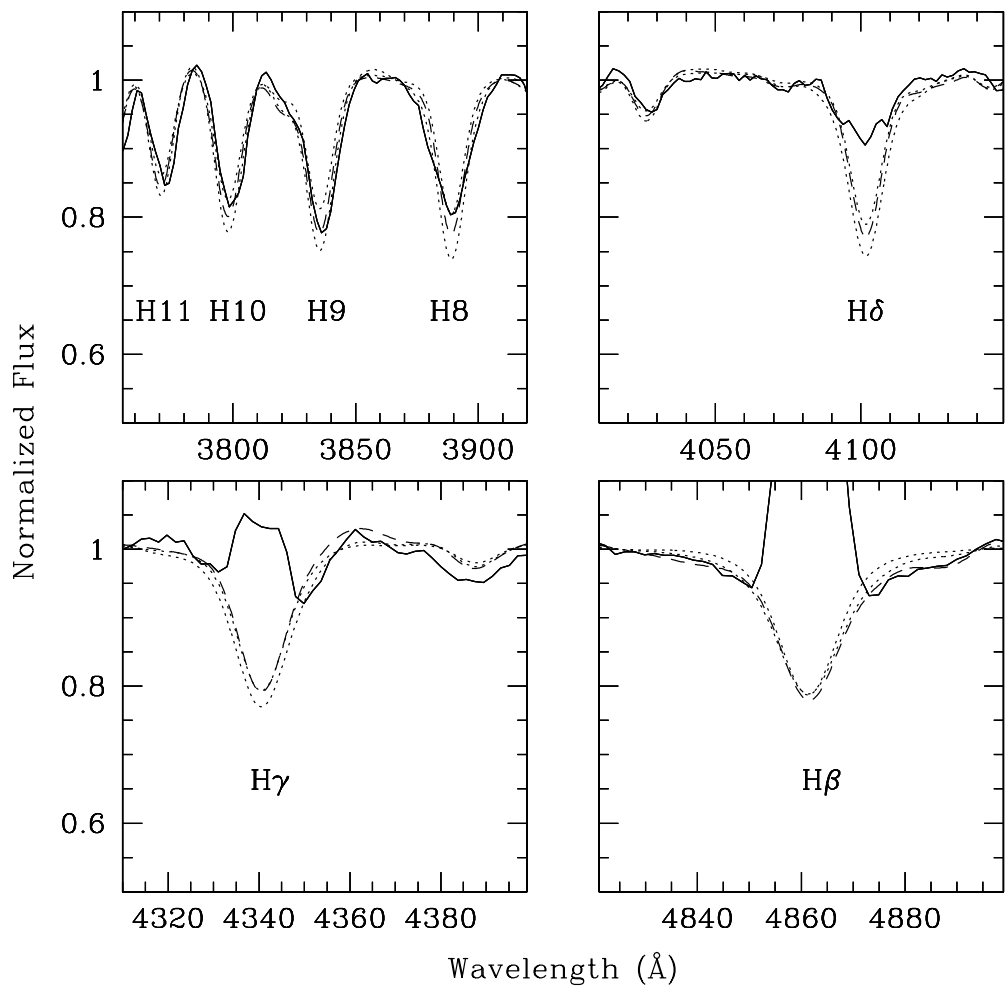}
\caption{The normalized spectra of the nuclear hot spot M83\,B are
compared to the models of \citet{gonzalez99} for ages of 4 (upper
dotted line), 6 (dashed line) and 8 (lower dotted line) Myr (Salpeter
IMF, instantaneous burst). The four panels show: (a) the higher-order
Balmer lines H8, H9, H10 and H11; (b) $H\delta$; (c) $H\gamma$; (d)
$H\beta$. The models have been degraded to the 10-\AA~resolution of
the data.\label{gonzales}}
\end{figure}

As mentioned in Section~2, the presence of the higher order Balmer
lines and He\,I lines in absorption in the optical spectra of the
circumnuclear \hii\/ regions is a signature of young stars of type O,
B and A, which contribute to most of the blue continuum.  These
features are seen in the integrated spectra of starbursts
(\citealt{vacca92}, \citealt{storchi95}) and of some EHRs
(\citealt{delgado00}). To estimate the ages of the hot spots from
these lines we have performed fits of the observed profiles of the
absorption components of the Balmer lines to the evolutionary
synthesis models of \citet{gonzalez99}. For the lower-order stellar
Balmer lines, often filled by nebular emission, we restricted our
analysis to the profiles in the line wings. An example of the
procedure is shown in Fig.~\ref{gonzales}, where we show the
normalized observed profiles of M83\,B in four different spectral
ranges, centered on $H\beta$, $H\gamma$, $H\delta$ and the next
higher-order Balmer lines. The spectra were rectified adopting the
continuum windows defined by \citet{gonzalez99b}. The instantaneous
burst, solar metallicity, Salpeter IMF models are shown for ages of 4,
6 and 8 Myr, and their spectral resolution has been degraded to match
the resolution of our data. Although the best fit is obtained with the
6 Myr model, the quality of the data does not allow us to put tight
constraints on the ages, even if values larger then 10 Myr are ruled
out.  We obtain similar results for all the remaining hot spots in
M83, except for M83\,A, for which the model fits suggest an age of 4 to
5 Myr, albeit with large uncertainty, since even the high-order Balmer
lines are filled with emission. This result is in overall agreement
with the estimate obtained from \ew.  Observations at higher spectral
resolution would allow us to put tighter constraints on the derived
ages. It is worthwile to mention that fits of similar quality can be
obtained with continuous star formation models, and discriminating
between the two star formation modes (burst vs.~continuous) is not
possible based on this kind of analysis alone. Similar conclusions
were reached by \citet{boker00}, who applied the same method to date
the nuclear star cluster in the starburst galaxy NGC~4449. Our
previous considerations on the strength of the He\,I lines, however,
rule out the possibility that in the single hot spots star formation
proceeds continuously.  The possibility remains that in the
circumnuclear region as a whole star formation is sustained for long
periods of time by the bar-driven gas inflows near the inner Lindblad
resonance. Star formation possibly extending for several hundreds of
Myr has in fact been suggested for the circumnuclear rings of NGC~5248
and NGC~1512 by \citet{maoz01}.

Our age estimates are in good agreement with other studies. From HST
broad-band imaging \citet{heap93} found several OB star clusters in
the nuclear region of M83.  A large number of these clusters is
visible in the STIS image shown in Fig.~\ref{m83stis.fig}, and we
identify them as the ionizing sources of our hot spot \hii\/
regions. \citet{heap93} derived ages of 2 to 6 Myr, based on the
dereddened cluster colors. This result is confirmed by the multi-band
HST imaging of the nuclear clusters obtained by \citet{harris01}, who
found a peak in the age distribution between 5 and 7 Myr.  The hot
spots studied here correspond to clusters with ages of 5 to 6 Myr,
with the exception of M83\,A, which contains clusters having ages
between 1 and 5 Myr.  Studies of the infrared $K$ band emission peaks
\citep{elmegreen98,thatte00} indicate ages younger than 10 Myr for the
main star forming regions in the nucleus of M83. Finally, based on IUE
spectra of NGC~3351, \citet{colina97} report ages of 4--5 Myr from
model comparisons of the measured Si\,{\sc iv}/C\,{\sc iv} equivalent
width ratio.

We also note that in M83 the hot spot ages are confined within a
rather limited time interval, suggesting a burst of star formation
about 5-7 Myr ago, and that very young ages ($<$\,3 Myr) are not
observed. This might also be the result of a selection effect or be
related to a minimum time required for the bright, optically selected
hot spots to burst out of their parental molecular clouds and become
visible, explaining while hot spots in M83, NGC~3351 and other
galaxies seem to posses roughly the same ages, as judged by their low
\ew.  The M83 images presented by
\citet{harris01} show that the regions occupied by our hot spots are
relatively free of dust lanes, probably as an effect of the star
forming episode itself, but the general distribution of the dust is
very complex, being possibly mixed with the gas, and we cannot
exclude the presence of highly extinct younger star forming regions.

\subsection{UV spectra}
\begin{figure}
\plotone{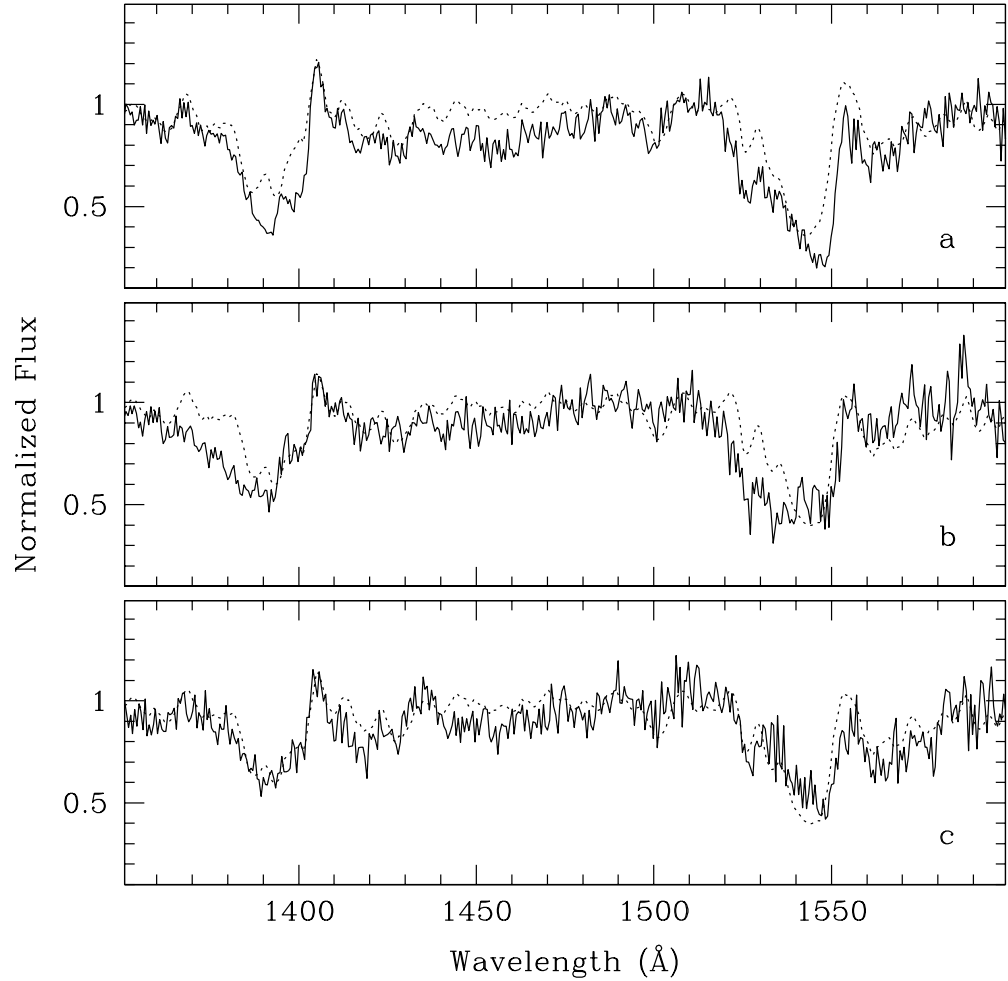}
\caption{Archival ultraviolet HST/STIS rectified spectra of three young clusters
located in the hot spot M83\,A. The dotted lines represent theoretical
Starburst99 models for solar metallicity, instantaneous bursts at ages
of 3.5, 4 and 4 Myr, from top to bottom (Salpeter IMF,
$M_{up}=100$~\msun).
\label{stis}}
\end{figure}

To further characterize the stellar population in the hot spots, we
show in Fig.~\ref{stis} a portion of the rectified vacuum-UV spectra
of three star clusters located in the circumnuclear ring of M83,
centered on the resonance Si\,{\sc iv}\w1400 and C\,{\sc iv}\w1550
lines.  These lines are among the most important stellar features
observed in the UV, being direct indicators of the presence of massive
stars, whose strong winds determine the distinctive blueshifted
absorption or P Cygni line profile in the UV resonance lines
(\citealt{walborn85}).  Several recent papers concentrating on the UV
spectra of starbursts in nearby galaxies observed with HST
(\citealt{gonzalez99c}, \citealt{johnson99,johnson00},
\citealt{origlia01}, \citealt{tremonti01}) have greatly 
contributed to our understanding of starburst activity in galaxies
(see \citealt{leitherer02} for a recent review).

Ultraviolet spectra from IUE and the Hopkins Ultraviolet Telescope
with apertures covering the whole starburst ring in M83 and NGC~3351
have been available for some time (e.g.~\/\citealt{bohlin83}).  The
large Si\,{\sc iv} and C\,{\sc iv} equivalent widths measured from
these spectra, both on the order of 10~\AA, are consistent with those
found in metal-rich environments (\citealt{storchi95},
\citealt{heckman98}).  The Space Telescope Imaging Spectrograph (STIS)
MAMA detector data in Fig.~\ref{stis} allow us to isolate some of the
individual compact star-forming clusters. The original data, obtained
with the G140L grating and the 52\arcsec$\times$2\arcsec\/ slit, were
retrieved from the STScI archive (program 8785, PI: Heap).  The
spectra in Fig.~\ref{stis} represent three clusters encircled in the
optical hot spot M83\,A.  The blueshifted absorption components of the
Si\,{\sc iv} and C\,{\sc iv} lines are a signature of the presence of
massive stars also in this high-metallicity environment.  The fact
that P Cygni profiles are present in both lines indicate the presence
of OB supergiant stars (\citealt{leitherer95b}), which in an
instantaneous star formation scenario appear approximately 3 Myr after
the burst.  The similar aperture geometry between our 2\arcsec\/ long
slit spectra and the STIS spectra, obtained with the slit centered on
the three clusters (whose dimensions are much smaller than the slit
size), helps in the comparison of the optical with the UV spectra,
where the three clusters are well separated in the spatial
direction. In Fig.~\ref{stis} synthetic spectra from the Starburst99
package (\citealt{leitherer99}) are displayed as a comparison. We show
only solar metallicity, instantaneous burst models calculated with a
Salpeter IMF with lower and upper mass limits $M_{low}$\,=\,1~\msun\/
and $M_{up}$\,=\,100~\msun, since for the estimated ages the UV
spectra are rather insensitive to the IMF parameters.

The age of the models shown in Fig.~\ref{stis} is 3.5, 4 and 4 Myr,
from top to bottom, while somewhat younger ages (by approximately 0.5
Myr) are inferred from the models which use the 2\,\zsun\/ stellar
tracks.  Part of the differences between the observed spectra and the
synthetic ones can be attributed to a metallicity mismatch between the
M83 clusters and the Starburst99 models, which were calculated
including UV stellar templates at slightly subsolar chemical
composition. A significant metallicity dependence is seen in the
strength of the UV lines when comparing stellar libraries at
0.25~\zsun\/ and \zsun\/ (\citealt{leitherer01}), so we can expect a
similar effect between the current models and supersolar abundance cluster
spectra.  An enhanced wind opacity in higher metallicity OB-type
supergiants leads to higher mass-loss rates, and in a possible
increase of the terminal velocity (\citealt{kud00}).  An increased
metal content would strengthen the photospheric lines, e.g.~\/the
Fe\,{\sc v} lines at 1360-1380 \AA, Si\,{\sc ii}\ww1526,1533 and
Fe\,{\sc iv}\w1533, which can contaminate the blue profiles of the
wind lines.  Blanketing from photospheric metallic lines (mostly iron
in different ionization stages) might also be responsible for the
depression of the continuum around 1450~\AA.

In summary, the strength of the observed resonance lines and the model
comparison provide evidence for the presence of massive stars and for
young ages in the clusters observed in the UV. In particular, the hot
spot M83\,A contains three clusters with similar ages ($\sim$4 Myr), a
result which is consistent with the optically-derived age. At this age
the Geneva stellar models predict the presence of 30-35~\msun\/
supergiant stars (B0\,I), which tells us that the IMF must extend at
least up to this mass.

\section{W-R features}

The study of W-R stars, the direct descendants of massive O stars, in
star forming regions plays an important role in constraining
population synthesis and evolutionary models of massive stars and
burst properties (age, star formation rate, IMF; \citealt{cervino94}, 
\citealt{meynet95}, SV98, \citealt{leitherer99}, \citealt{mashesse99}).
In the optical spectra of nearby starbursts the most commonly observed
feature of these stars is the broad `blue bump' around 4650~\AA\/,
which includes broad stellar emission lines such as the He\,II\w4686
and N\,III\w4640 from WN stars (\citealt{conti91}). At longer
wavelengths, the C\,III\w5696 and C\,IV\w5808 emission lines,
originating in WC stars, can also be observed, although their
detection is usually more difficult (\citealt{schaerer99}).  The
investigation of W-R features in metal-rich \hii\/ regions is of
particular interest, since the number of W-R stars relative to the O
star population is expected to increase with metallicity on the basis
of evolutionary models (\citealt{meynet95}), as a result of the
decrease of the minimum mass required for an evolving massive star to
develop the W-R phase (between 21\,\msun\/ and 32\,\msun\/ at high
metallicity, depending on mass loss rate, \citealt{maeder94}).
Observations of starbursts with high chemical composition (solar and
above) are consistent with this expectation (\citealt{guseva00},
\citealt{schaerer00}).

\begin{figure}
\plotone{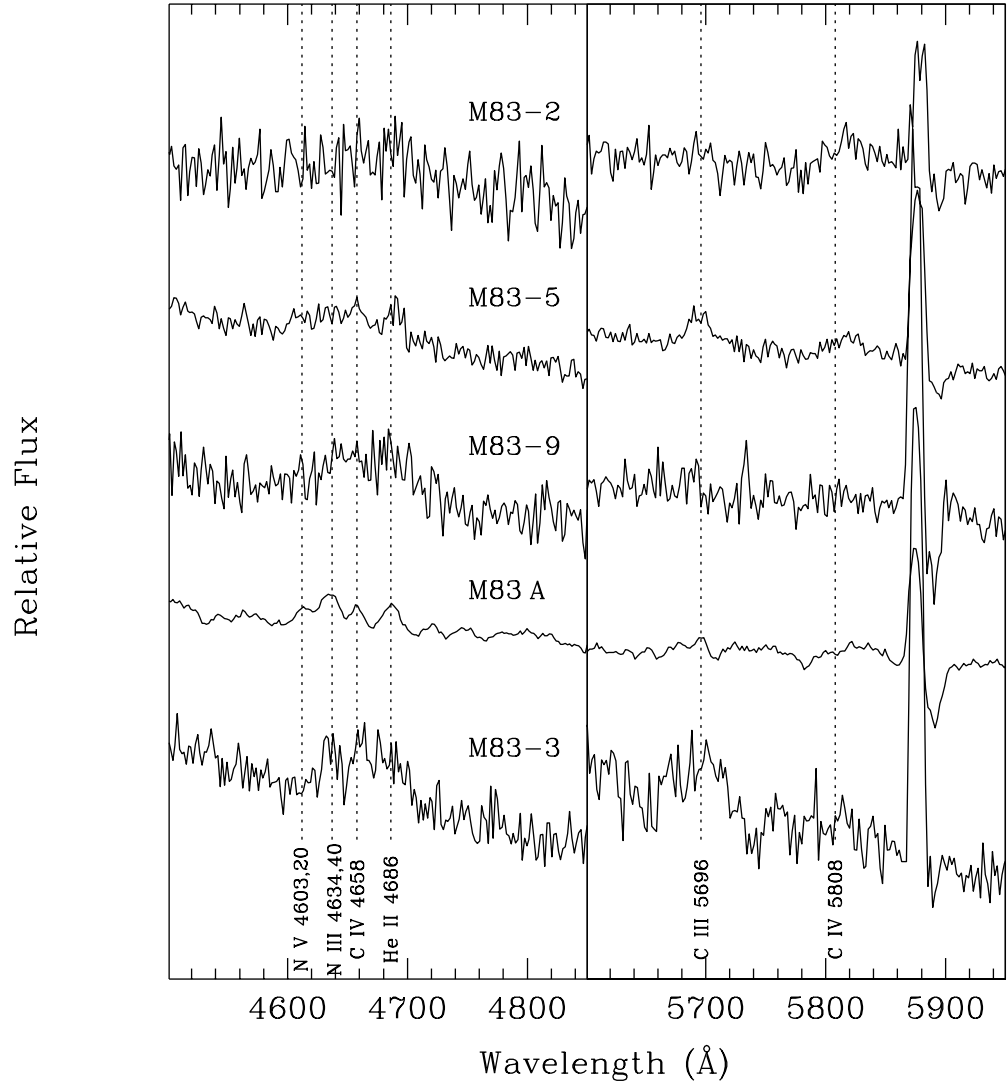}
\caption{Plot of the spectral ranges around W-R features in the blue
(left) and in the red (right) for the five objects with a W-R bump
detection in M83. The wavelengths of commonly observed stellar lines
from W-R stars are indicated.
\label{wr}}
\end{figure}

We report here on the possible detection of the W-R bump in five of
the \hii\/ regions we observed in M83. \hii\/ region M83-3 is one of
the three objects in which \citet{rosa86} detected the presence of W-R
stars, and corresponds to the position of supernova SN 1950b. The
remaining three disk \hii\/ regions (M83-2, 5, 9) and the hot spot
M83\,A would be new detections. The corresponding spectra around the
W-R features are shown in Fig.~\ref{wr}. For better clarity in the
plot the flux scale is different for all objects. 
The wavelength of several lines of stellar origin are indicated,
including N\,III\w4640, N\,V\w4604,4620 and the broad He\,II feature
at 4686\,\AA.  The right panel shows the region around 5800~\AA, where
features originating in WC stars are found, namely C\,III\w5696 and
C\,IV\w5808.  The detection of features in this wavelength range
remains uncertain for most objects, although in at least one case the
evidence for a detection of C\,III and C\,IV is rather convincing.

In the following we provide some brief comments about the five objects
whose spectra are displayed in Fig.~\ref{wr}. To estimate the dominant
W-R types we have adopted the classification scheme of \citet[see also \citealt{smith96} for the WN stars]{smith68}.

\smallskip
\noindent
{\em M83-2:} the S/N ratio of the spectrum is very low. We estimate an
extinction-corrected flux in the He\,II\w4686 line
$F(4686)=6.9\times10^{-16}$ erg~cm$^{-2}$~s$^{-1}$. At the adopted
distance of M83 this translates into a luminosity
$L(4686)=2.6\times10^{36}$ erg~s$^{-1}$. Here and in the following
discussion we adopt a typical WNL star luminosity
$L(4686)=1.6\times10^{36}$ erg~s$^{-1}$ (SV98), and
assume that all the flux in the He\,II\w4686 line originates in WN
stars. Our measurement is consistent with the presence of only one or
two WN stars in this \hii\/ region.

\noindent
{\em M83-5:} the similar intensity of N\,V\w4604,20, N\,III\w4640 and
He\,II\w4686 suggests an intermediate mean WN5 or WN6 type. From
$L(4686)=4.4\times10^{36}$ erg~s$^{-1}$ we estimate $\sim$3 WN stars.
This is the \hii\/ region where the detection of WC stars is more
convincing. Both C\,III\w5696 and C\,IV\w5805 are present. Their
intensity ratio is approximately equal to one, which suggests a late
W-R type, perhaps WC8.

\noindent
{\em M83-9:} the single components of the blend in the blue cannot be
separated. We estimate N\,III\,$\simeq$\,He\,II, which gives a He\,II
luminosity $L(4686)=1.5\times10^{37}$ erg~s$^{-1}$ (10 WN stars). The
N\,IV\w4058 line is detected, with an intensity relative to N\,III\w4640
of 0.2. These observations suggest a mean WN8 type. No WC line is
detected in the red.

\noindent
{\em M83\,A:} deblending of the different components of the blue bump
is less uncertain in this case.  N\,III\w4640 is somewhat stronger
than He\,II\w4686, and N\,V\,$\simeq$\,0.5~N\,III, suggesting a WNL
type, around WN7. From $L(4686)$ the estimated number of WN stars is
31.  C\,IV is probably not observed. The presence of absorption
features at \w5711 and \w5783 (as seen in the templates of
\citealt{bica88}) mimic an emission feature at the position of the red
bump. The presence of C\,III\w5696 is very uncertain.

\noindent
{\em M83-3:} N\,V is not visible, or is very faint. The relative
intensity of N\,III and He\,II again suggests a WN7 or WN8 mean type,
with $\sim$4 WN stars. Features from C\,III and C\,IV are not detected
at a significant level, although emission at the wavelength of
C\,III\w5696 might be present.

We have estimated the fractional contribution to the total ionizing
flux within the nebulae coming from the WN stars. The ionizing flux of
W-R stars was assumed to be $Q_0(WR)=10^{49}$ photons~s$^{-1}$,
irrespective of subtype (SV98). The nebular ionizing flux $Q_0(tot)$
was measured from the extinction-corrected luminosity of the $H\beta$
emission line, assuming the density-bounded case and no loss of
photons. The resulting $Q_0(WN)$/$Q_0(tot)$ varies between 0.13
(M83\,A) and 0.27 (M83-9). Under our assumptions, then, although WN
stars make a substantial contribution to the total ionizing flux, the
bulk of the ionizing photons comes from OB stars, unless several WC
stars are also present. The current observations do not allow us to put
credible constraints on this number based on the red bump. The
implications of the small number of W-R stars present in these \hii\/
regions will be further discussed in Sect.~7.3.

\section{Comparison with models and discussion}

In this section we compare the observed spectral features, namely
emission line ratios and equivalent widths, to the predictions of
recent evolutionary and photoionization models from the
literature. The recent advances in the synthesis of evolving stellar
populations at different metallicities, coupled with modern
photoionization codes, allow self-consistent investigations of the
properties of extragalactic \hii\/ regions, starbursts and \hii\/
galaxies.  The Starburst99 (\citealt{leitherer99}) and P\'{E}GASE
(\citealt{fioc97}) packages, in particular, have received a wide
acceptance, and have prompted the re-analysis of the bulk properties
of extragalactic emission line objects (\citealt{moy01},
\citealt{dopita00}, \citealt{kewley01}).  SV98 published a set
of evolutionary models tailored to the investigation of the massive
stars in young starbursts, based on the non-rotating Geneva stellar
evolutionary tracks (\citealt{meynet94}, and references therein), and
on the CoStar stellar atmosphere models
\citep{schaerer96, schaerer97}. New stellar tracks which
account for stellar rotation are available from the Geneva group
(\citealt{meynet00}, \citealt{maeder01}). Important changes in the
stellar evolution are predicted, however none of the currently
available evolutionary models include these newer tracks.  The
SV98 models were adopted by
\citet[hereafter SSL]{stasinska01b} in their analysis of \hii\/ galaxies.
As recognized in the latter work, an uncertain aspect of most of these
models is the treatment of the stellar atmospheres during the W-R
phase, based, in the case of the Starburst99 and the SV98 models, on
the pure helium \citet{schmutz92} models, which do not include
metallicity effects on the outcoming ionizing flux. The correct
treatment of metal blanketing is likely to introduce important changes
in the predicted ionizing flux of stellar populations during the W-R
phase (\citealt{crowther99b}, \citealt{crowther99}, BKG,
\citealt{smith02}).  

Further assumptions concern the mass-loss recipe adopted for the
stellar evolutionary tracks. This choice has important repercussions
on several aspects of massive star evolution. Of interest here are not
only the duration of the W-R phase and the minimum mass required for
stars entering this phase (increasing and decreasing, respectively,
with mass-loss rate), but also the stellar extreme-UV spectrum.
Recent population synthesis models often adopt the Geneva stellar
tracks calculated with an {\em enhanced} mass-loss rate, i.e. twice
the values given by \citet{dejager88} throughout the HR diagram
(excluding W-R stars), with a metallicity scaling derived from stellar
wind models, and $\dot{M}=8\times10^{-5}$ \msun/yr for WNL star
(\citealt{meynet94}). For WNE and WC stars the mass-loss rates are
kept fixed at the same values as in the {\em standard} case
(\citealt{schaller92}).  While empirical results, such as the W-R/O
number ratio in regions of constant star formation, favor enhanced
mass-loss rates in the pre-WR stage (\citealt{maeder94}), the actual
mass-loss rates measured for WNL stars are lower by at least a factor
of two with respect to the enhanced rates assumed in the Geneva tracks
(\citealt{leitherer97}; \citealt{nugis00}). These rates have been
considerably lowered (by a factor 2-3) in the newer tracks including
rotation (\citealt{meynet00}).

One important aspect of mass-loss rate which is relevant for the
following discussion is the fact that it controls the EUV flux (below
the He~II ionization edge at $\lambda=228$~\AA) in massive stars with
expanding atmospheres. The flux can be either enhanced or decreased,
depending on the wind density, by its effect on the depopulation of
the He~II ground state (\citealt{gabler89}). \citet[see also
\citealt{crowther99b}]{schmutz92} show that the presence of
significant flux above 4 Ryd depends on relatively transparent (low
$\dot{M}$) W-R winds.

The evolutionary models considered in the remainder of this section
are summarized in Table~6, where we indicate their main ingredients:
atmospheres adopted during different evolutionary phases (main
sequence and W-R), stellar tracks and mass-loss rate recipe. The
latter is expressed in terms of the mass-loss rate value adopted by
the Geneva group (1\,X and 2\,X for standard and enhanced mass-loss
rate, respectively).  A recent comparison of the different codes,
focussing on the W-R population synthesis, is given by
\citet{leitherer99iau}.  Most codes adopt the Geneva stellar tracks
(except for the Padova tracks in the case of
P\'{E}GASE). Different combination of atmosphere models are used,
including the more recent CoStar fluxes for O stars in the SV98 and the
\citet{cervino01b} models, and the
\citet{kurucz92} and \citet{lejeune97} atmospheres for the remaining
stars across the HR diagram. For stars in the W-R stage, either
specific models such as those of \citet{schmutz92}, or the NLTE hot
star models of \citet{mihalas72} or \citet{clegg87}, are used.

\subsection{Abundances}
\begin{figure}
\plotone{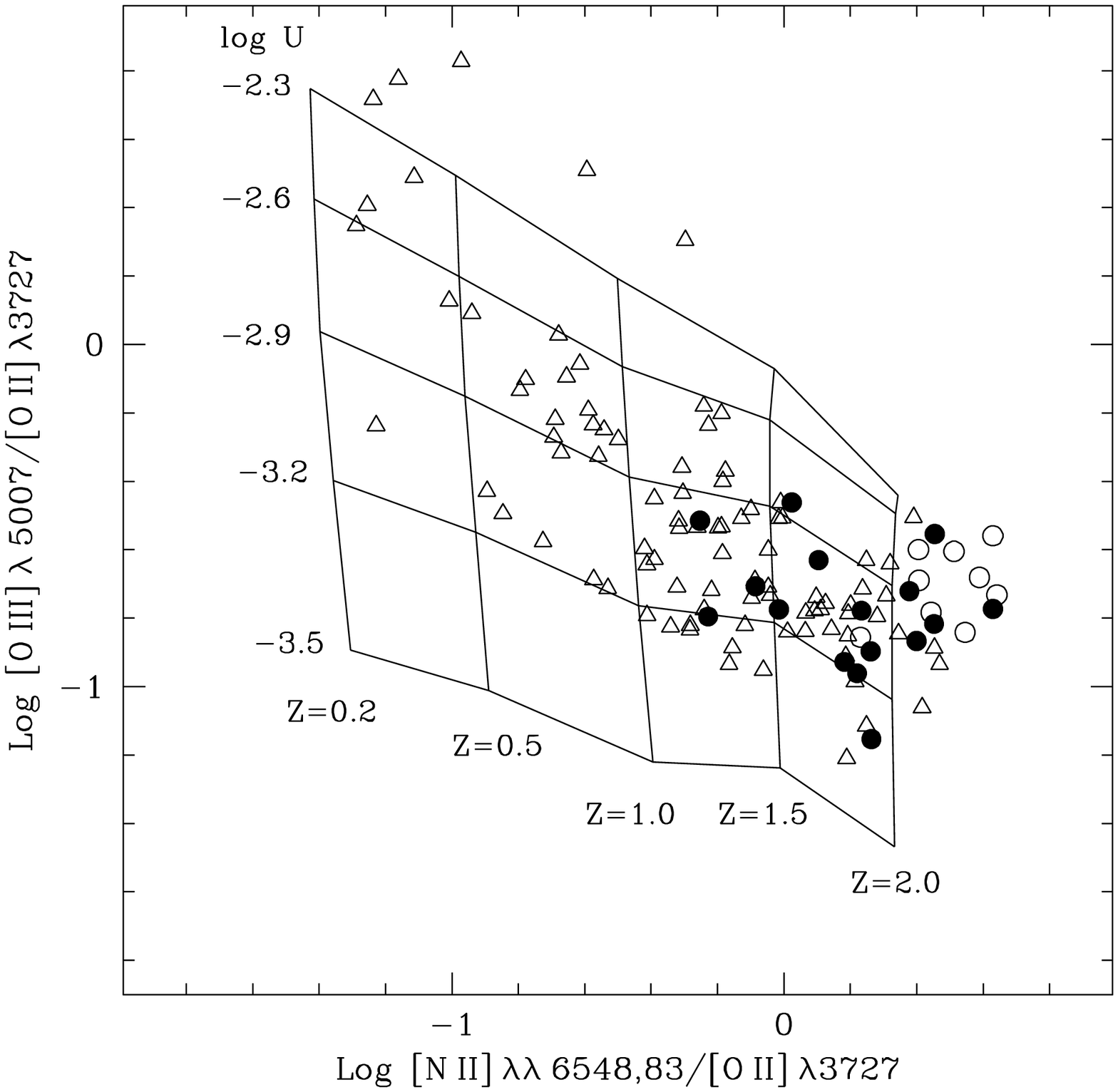}
\caption{Diagnostic diagram relating the abundance-sensitive line
ratio [N\,II]\ww6548,6583/[O\,II]\w3727 to the ionization
parameter-sensitive [O\,III]\w5007/[O\,II]\w3727 line ratio. The model
grid from \citet{dopita00} is superposed on the representative points
for the objects in the current sample (circles) and for the
\hii\/ regions studied by BKG (triangles). The theoretical abundances (Z)
and ionization parameter (U) values are indicated.
\label{dopita}}
\end{figure}

\citet{dopita00} have recalibrated the EHR sequence with models based
on both the P\'{E}GASE and Starburst99 predictions for the spectral energy
distribution of zero-age instantaneous bursts. These were used as
input for photoionization models which included a self-consistent
treatment of dust effects on the nebular spectra. Their conclusion is
that the bulk properties of EHRs can be satisfactorily explained by
young ($<$\,2 Myr) models, and proposed the use of the [O\,III]/[O\,II]
versus [N\,II]/[O\,II] diagnostic diagram to estimate both the chemical
abundance and the ionization parameter. In Fig.~\ref{dopita} we plot their
models, based on the P\'{E}GASE ionizing fluxes, on top of our new data,
together with the BKG sample. According to this diagram, and under the
assumption that these \hii\/ regions are very young, the metallicity
of our new sample is supersolar, with several objects, including the
hot spots, having abundances well above twice the solar value. The
apparently large discrepancy with the metallicities derived previously
from the strong line method underlines the difficulty of abundance
determinations in this $O/H$ regime. However, the [N\,II]/[O\,II]
depends on both $T_e$ (this ratio being larger at the lower
temperatures found in metal-rich nebulae) and N/O ratio. A different
recipe in accounting for how this ratio depends on $O/H$ would lead to
different [N\,II]/[O\,II] line ratios. This is seen when the
\citet{stasinska01b} models are considered, which adopt a milder 
dependence of N/O upon metallicity than the \citet{dopita00}
models. This is the likely explanation for their generally small
predicted [N\,II]/[O\,II] ratios at high abundance which, even in the
2\,\zsun\/ case, cannot account for the large values
([N\,II]/[O\,II]~$\simeq$~3) observed in metal-rich
\hii\/ nebulae. An increasing N/O ratio with time would also shift the
model grid to the right in Fig.~\ref{dopita}, leading to smaller
metallicities estimates with respect to the zero-age case.

How does the abundance calibration adopted for this work, based on
nebular models and zero-age ionizing cluster models (see Section~3.3)
compare with recent photoionization models including time evolution?
SSL have pointed out that at high metallicity the sensitivity of the
optical [O\,III] lines to the balance of energy gains and losses makes
the strong line method of abundance determination, together with the
auroral-line based method, highly uncertain, and as the ionizing
clusters evolve the empirical abundances will correspondingly show an
important variation.  In order to estimate the effect on our abundance
scale, we have used the
\citet{kobulnicky99} analytical expression of the McGaugh models
(upper branch only of the $R_{23}$-$O/H$ relation) and applied it to
the time-evolving fluxes predicted by the SSL models.  We have used
their reference IMF models (Salpeter slope, $M_{up}$\,=\,100~\msun),
with total cluster mass of $10^6~M_\odot$, to calculate semi-empirical
abundances from the model [O\,II] and [O\,III] lines. The results are
little affected by changes in the total mass, and are summarized in
Table~4 for instantaneous burst models at metallicities \zsun\/ and
2\,\zsun.  The resulting semi-empirical abundances indeed show a
variation with time, starting at approximately the theoretical input
value, and then departing from it already 2 Myr after the initial
burst. The departures can be quite large, and act in the sense of
either over- or under-estimating the input model abundance, depending
on the evolutionary status.  For ages smaller than 6 Myr, however, the
scatter is confined within $\sim$0.15 dex of the input abundance.
This comparison indicates that, while it could be very difficult to
assign accurate abundances to metal-rich nebulae, especially at
abundances well above the solar value, we can confirm the supersolar
abundances of most objects in the present sample of EHRs in M83,
NGC~3351 and NGC~6384.

\subsection{Ages}
\begin{figure}
\plotone{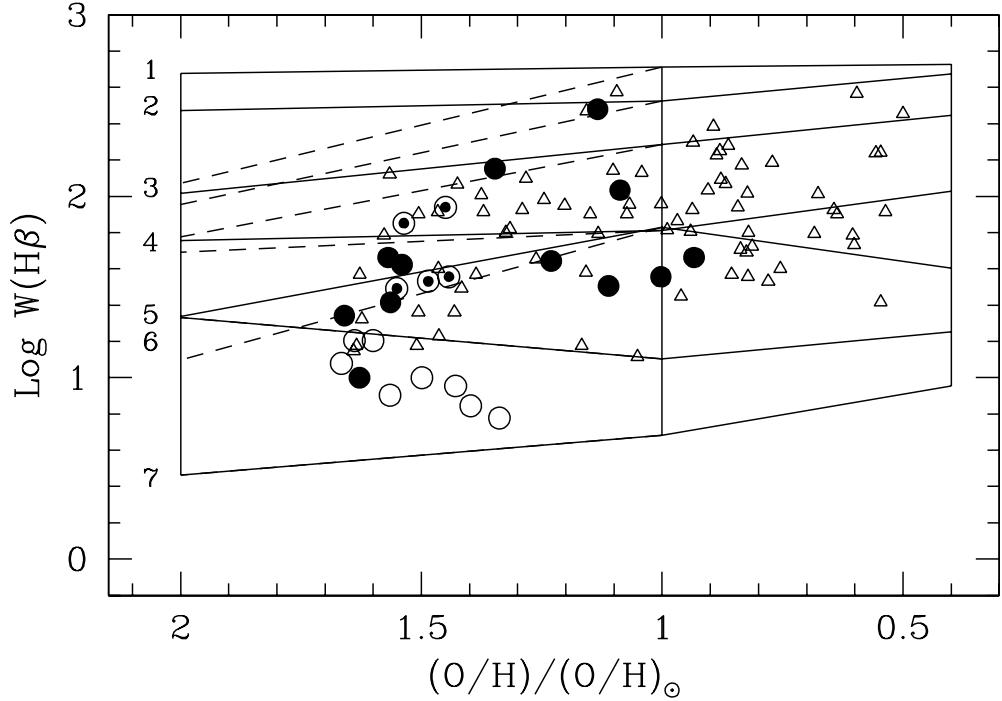}
\caption{Metallicity dependence (relative to solar) of \ew\/ for the
objects measured in M83, NGC~3351 and NGC~6384 (circles), and for the
\hii\/ regions studied by BKG (triangles). The five \hii\/ regions in
M83 with a W-R blue bump detection are indicated by the semi-filled
circles. The lines connect the values from the photoionization models
of SSL at three different metallicities (0.4\,\zsun,
\zsun\/ and 2\,\zsun), and at different cluster ages, from 1 to 7 Myr
(at 1 Myr age intervals, indicated on the left), calculated for a
Salpeter IMF and $M_{up}=120$~\msun. The dashed line connect the solar
abundance models to models at twice this abundance, calculated for
$M_{up}=30$~\msun.
\label{ew_oh}}
\end{figure}

The age of a starburst is an additional, fundamental property which,
in principle, can be derived from the observed spectrum. The $H\beta$
equivalent width, \ew, is widely adopted as a starburst chronometer
(\citealt{copetti86}), and at metallicities below solar also
$W$([O\,III]\w5007) could be employed (\citealt{stasinska96}). However,
difficulties in the use of \ew\/ can arise if a significant older
underlying stellar population is present, as shown by
\citet{raimann00}. As a consequence, ages determined from
\ew\/, as done in Section~5, would in general be overestimated. In
Fig.~\ref{ew_oh} we plot \ew\/ vs.~$O/H$ for our metal-rich \hii\/
region data, supplemented by the BKG data. The SSL models for a
Salpeter IMF and $M_{up}$\,=\,120~\msun\/ are plotted for starburst ages
varying from 1 to 7 Myr. The dashed lines connect the solar abundance
predictions with those for a $M_{up}$\,=\,30~\msun\/ IMF at twice solar
abundance. The well-known lack of \hii\/ regions with \ew\/
values in excess of 500~\AA\/, as predicted by photoionization models,
is apparent at all abundances. This does not necessarily imply the
absence of very young objects, and alternative explanations have
been proposed. These include uncertainties in the evolutionary models
(e.g.~\/overprediction of ionizing photons, see~\citealt{mashesse99}) as well as effects of dilution by underlying stellar
populations (\citealt{raimann00}), dust extinction affecting stars and
gas in different proportions (\citealt{mashesse99},
\citealt{schaerer00}), and leakage of ionizing photons (SSL).
The absence of a proper zero age main sequence for massive stars and
the considerable time, on the order of 2 Myr, they spend embedded in
their parental clouds, as proposed by \citet{bernasconi96}, does not
seem to reduce \ew\/ by a suffient amount to reconcile it with the
observational data (\citealt{schaerer00}).  Although noted and
addresses in several papers, this discrepancy therefore still lacks a
unique and clear explanation.

In the high-abundance bins of Fig.~\ref{ew_oh} we notice several
objects, including the hot spots, with very low \ew. This diagram
would imply that the bulk of the sample has ages between 3 and 5 Myr,
with a number of older objects at high metallicity. It is however
possible that an upward correction in \ew\/ should be considered, in
order to account for an underlying older population, which is more
likely to be present at high abundance and in the nuclear hot
spots. This correction would lower the average age of the \hii\/
regions in the sample, but it is not possible in the current analysis
to exactly quantify this effect. The detection of the W-R bump in five
objects of the present sample, marked by the special symbols in
Fig.~\ref{ew_oh}, does not impose very tight constraints. According to
the SV98 models, W-R star signatures should be visible in the spectrum
of starbursts having ages in the range approximately between 2 and 7
Myr in the case of \hii\/ regions at solar abundance or above.  The
points corresponding to the W-R bump detections in Fig.~\ref{ew_oh}
are all in agreement with this prediction.  In Sect.~5, from the fit
to the optical absorption line profiles and the UV spectra, we
estimated for M83-A an age of 4-5 Myr, close to the age derived from
\ew.  We also note that, except for one object (NGC~3351-1), all of the low
\ew\/ points in Fig.~\ref{ew_oh} are hot spots. The absorption line
profile fits to their spectra in Sect.~5 suggested ages between 5 and
6 Myr, although the uncertainty can be considerable. An upward
correction in \ew\/ by $\sim$0.5 dex would be required to reconcile
these ages with those estimated from \ew.

\begin{figure}
\plotone{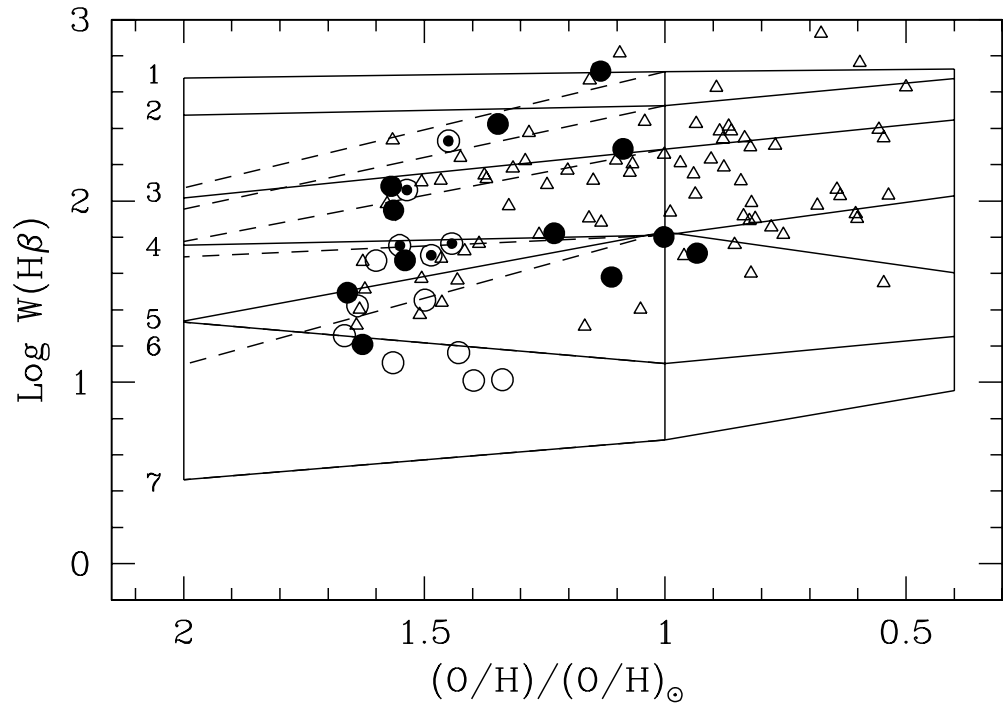}
\caption{As in Fig.~\ref{ew_oh}, but correcting \ew\/ for the
different amounts of extinction affecting the gas and the stars,
following \citet{calzetti01}.
\label{ew_oh2}}
\end{figure}

In principle, we have already accounted for an underlying stellar
population, since it resulted from our calculation of the reddening
(Sect.~2), based on the simultaneous convergence of the observed
$H\alpha$/$H\beta$ and $H\gamma$/$H\beta$ ratios to the theoretical
values.  However, \citet{mashesse99} found that values of $W_{abs}$
larger than the typical 1-2~\AA\/ measured in this way, and up to
$\sim$\,5~\AA, are required in some extragalactic star forming regions
to match the \ebv\/ values derived from the Balmer decrement to
those derived from the UV continuum. Indeed, the evolutionary models
of \citet{gonzalez99} predict $W_{abs}$($H\beta$)\,$\simeq$\,5~\AA\/
at $t=5$ Myr for an instantaneous burst at solar metallicity.
\citet{mashesse99} also found that the discrepancy between the optical
(Balmer lines) and UV (continuum) \ebv's, already mentioned
previously, is higher for the older \hii\/ regions, and at high
metallicity.  This can have an important impact on \ew, in that it
measures the ratio of nebular emission lines (higher reddening) to the
stellar continuum (lower reddening). The likely explanation for this
effect is that powerful stellar winds and supernova explosions sweep
the dust out of the ionizing stellar clusters into filaments within
the \hii\/ regions.
The difference measured by \citet{mashesse99} between stellar and
nebular extinction at older ages would lead to underestimate \ew\/ by
approximately a factor of 2. We argue that this might explain, at
least partially, the apparent older ages at high oxygen abundance in
Fig.~\ref{ew_oh}, and reconcile the ages determined via line profile
fits and \ew. As a further test, we plot in Fig.~\ref{ew_oh2} the same
quantities as in Fig.~\ref{ew_oh}, where we have applied a correction
to \ew\/ following \citet[see also \citealt{calzetti94}]{calzetti01}:
log\,[\ew$_{obs}$/\ew$_0$]\,=\,$-0.64$\,\ebv$_{gas}$. The resulting
\ew\/ values indeed show how the effect of differential reddening 
produces somewhat younger ages overall, and partially removes the
difference in \ew\/ between hot spots and disk \hii\/ regions.

We finally note that in the age range which includes most of the
\hii\/ regions in our sample the effects of an IMF depleted in massive
stars are secondary, as judged from the corresponding models in
Fig.~\ref{ew_oh}, and we can conclude that such a truncated mass
function is not required to explain the distribution of \ew\/ with
metallicity.

\subsection{W-R features}
The detection of W-R features in some of the \hii\/ regions of the
present sample (Sect.~6) allows us to derive some important properties
of the massive star content direcly from stellar features, rather
than relying on the more indirect signatures represented by the
nebular emission lines. The study of these features at high
metallicity is of particular importance, because it helps us
constraining current evolutionary synthesis models of massive stellar
populations in supersolar abundance environments.  So far only in a
handful of extragalactic metal-rich
\hii\/ regions these stars have been detected and analyzed
quantitatively (\citealt{schaerer99}, \citealt{schaerer00}).  Our
observations do not reach a sufficient S/N ratio for an accurate
analysis, but we can still infer some important properties of the
ionizing stellar content of those clusters containing W-R stars, by
comparing our observed line intensities and equivalent widths with the
model predictions (mainly from SV98).  We complement our sample with
six additional metal-rich \hii\/ regions from the work of BKG, which
show the presence of the W-R bump in their spectra (this analysis was
not included in the BKG paper). Of these, only one object (M31-5 using
the BKG nomenclature) shows a significantly strong red bump at
5808~\AA.

In the following analysis we will make a series of assumptions. First
of all, the extinction affecting the ionizing stars will be taken to
be equal to the one measured for the gas from the Balmer decrement. As
mentioned in the previous section, there are indications that the
stars suffer less extinction than the gas
(e.g.~\/\citealt{mashesse99}), and this would result in smaller
numbers of W-R stars as estimated from the intensity of the blue
bump. We will also rely on \ew\/ as an age indicator, and assume that
all Lyman continuum photons contribute to the ionization of the gas,
and that during the observations our slit included most of the
emission from the ionized gas, i.e.~we do not account for possible
aperture effects. We refer the reader to the work of
\citet{schaerer99} for a discussion on the impact of these
assumptions.

\begin{figure}
\plotone{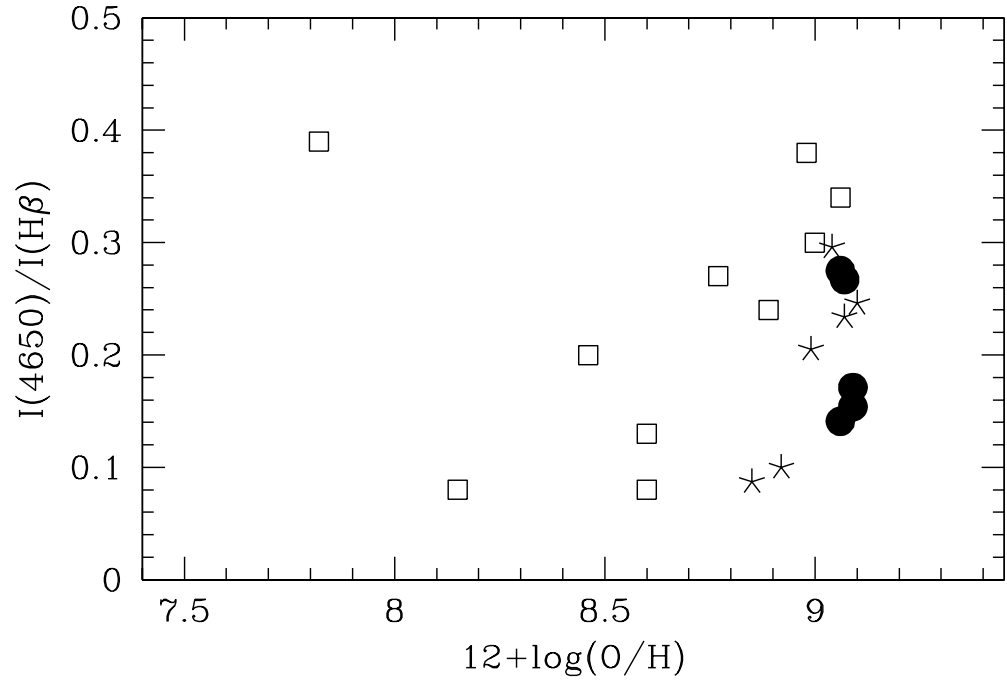}
\caption{Metallicity dependence of the intensity of the W-R 4650~\AA\/
blue bump relative to $H\beta$. The filled circles represent the five
\hii\/ regions in M83 where the W-R bump has been detected. Star
symbols are used for six additional objects drawn from the BKG sample
of extragalactic \hii\/ regions. W-R galaxies from the
\citet{schaerer00} work are represented by the open squares. The
abundances for the latter were recalculated, based on the strong line
method calibration adopted in the current work.
\label{wr_oh}}
\end{figure}

Given the uncertainties involved in the measurement of the individual
lines contributing to the W-R blue bump in our data, we will only
consider their combined fluxes. Apparently, no contamination from
nebular lines is present. The blue bump intensities relative to
$H\beta$ and their equivalent widths are summarized in Table~5 for the
five \hii\/ regions in M83 described in Section~6, and for the
additional EHRs from the BKG work (the identification refers to their
Table~3).  Uncertainties are of the order of 10-20 percent in the
intensities, and 1-2~\AA\/ in the equivalent widths. The
I(4650)/I($H\beta$) intensity ratio as a function of oxygen abundance
is shown in Fig.~\ref{wr_oh} for our objects, together with datapoints
for W-R galaxies taken from
\citet{schaerer00}. For the latter dataset we have, for consistency,
estimated the oxygen abundance using our adopted strong line method
calibration from the published strength of the [O\,II] and [O\,III]
emission lines. There is a general agreement with the known trend of
increasing I(4650)/I($H\beta$) with abundance, although the \hii\/
regions tend to have a weaker blue bump at a given abundance compared
to the brighter starburst galaxies.  In this plot we cannot exclude
systematic differences arising from observational biases.

\begin{figure}
\plotone{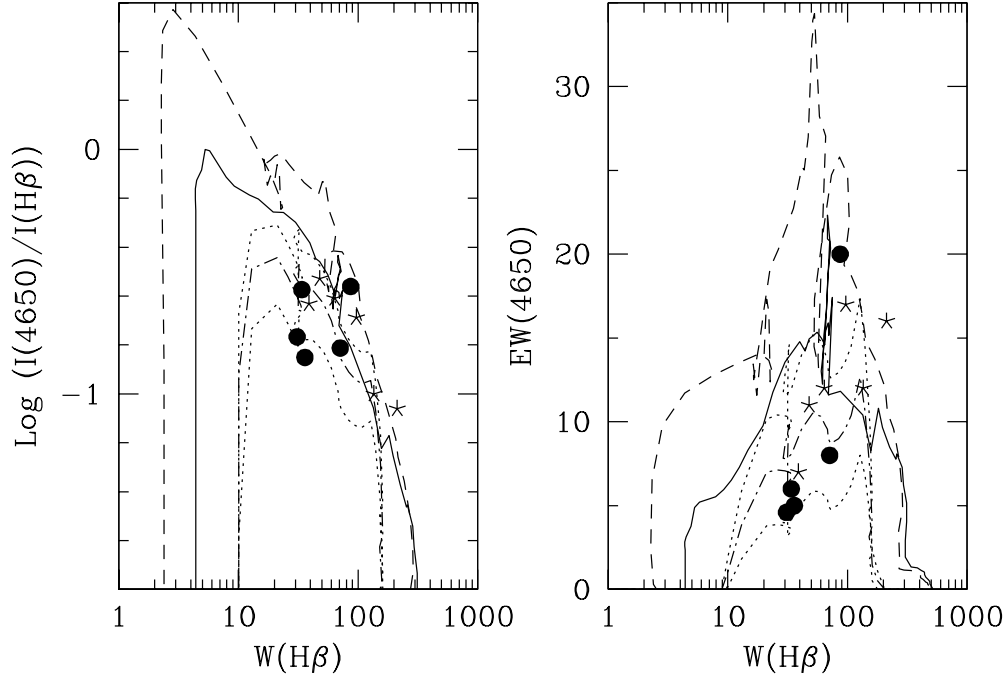}
\caption{Observed quantities of the W-R blue bump. {\em Left:} Bump
intensity relative to $H\beta$ as a function of \ew. The
SV98 model predictions for an instantaneous burst having
a Salpeter IMF with $M_{up}=120$~\msun\/ are shown (\zsun: full line;
2\,\zsun: dashed line, enhanced mass loss stellar tracks used). The
filled circles represent the five \hii\/ regions in M83 where the W-R
bump has been detected. Star symbols are used for additional
objects drawn from the BKG sample. The analytical prediction from
\citet{cervino01b} is plotted with a dot-dashed line (standard mass
loss). The 90\% confidence limits are shown by the dotted lines.  {\em
Right:} Equivalent width of the blue bump as a function of
\ew. Same symbols and models as in the previous plot.
\label{sv98}}
\end{figure}

Our data are compared in Fig.~\ref{sv98} to the models of SV98 for the
metallicities appropriate for our sample ($Z=Z_\odot$, full line, and
$Z=2~Z_\odot$, dashed line). The model IMF has a Salpeter slope with
$M_{up}=120~M_\odot$.  It can be seen that for half of the objects
there is good agreement between the observed quantities (both
I(4650)/I($H\beta$) and $W(4650)$) and those synthesized for an
instantaneous burst of star formation. The remaining ones have low
$W(4650)$ and/or I(4650)/I($H\beta$) values at a given \ew\/ if
compared with these models, and are all concentrated at the lower end
of the \ew\/ range spanned by our sample of objects.  A discrepancy in
the same direction was obtained for the five metal-rich W-R galaxies
studied by \citet{schaerer00}, who resolved the inconsistency by
considering models for bursts extending for 4-10 Myr, instead of
instantaneous ones. Under the same assumption, the burst duration for
the objects in our sample would have to be longer, because of the
lower blue bump intensities and equivalent widths of the discrepant
objects.  However, we find such mode of star formation less plausible
in the case of disk \hii\/ regions than for starburst galaxies,
which compose the objects considered in the
\citet{schaerer00} work, and which are likely the 
superposition of several ionizing clusters in a kpc-scale star forming
region. We do not detect in our spectra the signatures of older
generations of stars, except in the special case of the hot spot
M83\,A, where we attributed such features to the old bulge
population. More sensitive observations would be needed to confirm
this, however.  Observational biases, and in particular aperture
effects, would make the comparison with the models even worse, by
lowering the I(4650)/I($H\beta$) ratio (while $W(4650)$ would remain
unaffected). The same is true if we considered an extinction for the
stars lower than for the gas. Loss or leakage of ionizing photons,
e.g.~\/from dust absorption or density-boundedness, could not explain
the discrepancy in both plots (\ew\/ and I(4650)/I($H\beta$) would be
affected, but $W(4650)$ would not). \citet{schaerer00} also showed
that IMF parameters far from standard are unlikely to occur, in
particular upper mass cutoffs as low as 30~$M_\odot$ are ruled out.  A
more moderate decrease in the upper mass limit of the Salpeter IMF
from 120 to 60~\msun\/ cannot be excluded with the same degree of
confidence, and would be consistent with the low-EW\/
objects. However, if the measured \ew\/ values in excess of 100~\AA\/
are real, and not an effect of wrongly placed low continuum levels,
this solution would not be acceptable, either.

A further hypothesis follows from noting that the discrepant objects
in Fig.~\ref{sv98} are those with the lowest \ew\/ and $W(4650)$. We
can postulate that the low-\ew\/ end of the diagrams is not populated
because at high metallicity the W-R phase has a shorter duration than
predicted by the models. In fact, if the end of the W-R phase occurred
at $\sim$6 Myr rather than $\sim$7 Myr as predicted by the SV98 models
at high metallicity, the abrupt decrease of both I(4650)/I($H\beta$)
and $W(4650)$ would take place at higher values of \ew\/ (around
30~\AA), thus removing the discrepancy.  Although this solution might
seem perhaps justified by the uncertainties in our current knowledge
of the W-R phase at high metallicity, its verification relies on the
observation of larger samples of metal-rich \hii\/ regions. The
measurement of W-R bump emission in \hii\/ regions having
\ew\,$<$\,10~\AA\/ would refute this hypothesis. Such objects would
likely be faint because of their old age, and subject to considerable
measurement uncertainties.  The sample of
\citet[see \citealt{schaerer99b}]{arnault89} appears to contain a
handful of such objects.  We must also mention the fact that the newer
stellar tracks from the Geneva group including stellar rotation would
predict longer W-R lifetimes than the non-rotating models, and not the
opposite.

One effect which could, at least in principle, solve the discrepancy
noted above is related to the stochastic nature of the IMF that occurs
for small cluster masses.  As discussed by \citet{cervino00},
observables like \ew\/ and the intensity of the W-R bump are affected
by significant fluctuations when ionizing clusters of total mass
$10^3-10^4~M_\odot$ (accounting for stars in the 2-120~\msun\/ mass
range) are synthesized with Montecarlo-generated IMFs rather than with
analytical functions. In these cases a meaningful comparison with the
observational data should include the study of these
fluctuations. These effects can be quite important in the study of the
emission line properties of EHRs. In our sample the cluster masses
corresponding to the observed ionizing fluxes are in fact on the order
of $10^4~M_\odot$, except for M83\,A, which is one order of magnitude
more massive (similar moderate masses have been measured for the young
clusters in the M83 nuclear region by
\citealt{harris01}). \citet{cervino01b} have shown that during the
W-R--rich phase of an evolving starburst, clusters having total masses
smaller than $\sim$\,$10^6$~\msun\/ ($\sim$\,$10^5$~\msun) will have
an uncertainty in the number of He$^+$ (He$^0$) ionizing photons
larger than 10\%.  We have also seen in Sect.~6 that the number of W-R
stars contained in our sample \hii\/ regions is very small, therefore
small number statistics certainly plays an important role in the
interpretation of the diagrams presented in Fig.~\ref{sv98}.

From the \citet{cervino01b} work we have estimated the fluctuations in
the quantities plotted in Fig.~\ref{sv98}, by assuming a cluster mass
of $3\times 10^4$~\msun. The dash-dotted lines show the analytical
predictions, while the dotted lines give the upper and lower limits
corresponding to the 90\% confidence level, for the solar abundance
case. These models were generated with the Geneva stellar tracks and
the {\em standard} mass-loss rate, which makes the analytical solution
to lie below the corresponding prediction from SV98, since less W-R
stars are produced with a lower mass loss.  Despite this difference, the range in which we
expect to find objects roughly corresponding to those in our sample is
compatible with the location of the datapoints in both diagrams. This
solution is particularly attractive, since we do not have to postulate
changes in the IMF or very extended periods of star formation in order
to explain the diagrams in Fig.~\ref{sv98}. Future work on the
comparison of improved evolutionary codes with larger samples of
metal-rich \hii\/ regions should shed new light on this
interpretation.

\subsection{Equivalent effective temperatures}
The equivalent effective temperature of a stellar population, \teff,
defined as the temperature of a single hot star having the same
$Q_{He}/Q_H$ ratio of He\,I and H ionizing photons as the ensemble of
ionizing stars, is a useful diagnostic of EHRs.  The decrease of the
fractional abundance of ionized helium with increasing oxygen
abundance has led \citet{panagia00} to argue in favor of a decreased
upper mass limit for the IMF at high metallicity (as in BKG), or an
increase in the absorption of ionizing photons from the circumstellar
material.  We have shown in the diagrams in Section~4, in particular
the one relating the He\,I\,\w5876/$H\beta$ emission line ratio with
abundance, an empirical trend of the ionizing cluster
\teff\/ with metallicity, reaching \teff$\;\simeq\;$33-35$\times 10^3~K$
at supersolar abundances.  Theoretically, such a trend is expected
from the effects of metal abundance on the stellar temperatures, and
\citet{mcgaugh91} showed, using zero-age models of ionizing clusters,
how this would explain the decreasing hardness of the ionizing spectra
with metallicity.  However, \teff\/ is expected to decrease also as
the ionizing clusters evolve with time, as a result of the progressive
disappearance of the hottest and most massive stars
(\citealt{cervino94}). At intermediate ages (3-6 Myr) the appearance
of W-R stars could, as a result of high effective temperatures,
disrupt this trend by increasing the hardness of the cluster ionizing
spectra (\citealt{vargas95}). We attempt now to use our observations
to verify the predictions of current evolutionary models regarding
this particular aspect. In our previous work (BKG) we argued about a
possible depletion of massive stars in the IMF of metal-rich
\hii\/ regions. This argument followed from the result that the ages
of the \hii\/ regions in our sample were preferentially restricted to
the first 3 Myr after the onset of star formation. We reached this
conclusion by comparing the data with photoionization models based on
the cluster SEDs of \citet{leitherer95}. BKG proposed that, unless the
hardness of the ionizing spectra predicted by the models was
overestimated, an agreement with the data was found only for young
ages. In view of the progress in evolutionary synthesis modeling and
the availability of improved stellar tracks with respect to those used
in the \citet{leitherer95} work (the \citealt{maeder90} tracks),
leading in particular to shorter phases of enhanced $Q_{He^+}/Q_H$
from W-R stars at high metallicity, we re-analyze here the BKG data,
together with our new data, regarding inferences on the IMF from the
nebular lines. In fact, we are not forced to assume preferentially
young ages for the EHR in our samples, and have used the \ew\/
throughout this paper as a reliable time indicator. Lower \teff's can
thus be obtained also as a result of aging.

\begin{figure}
\epsscale{.9}
\plotone{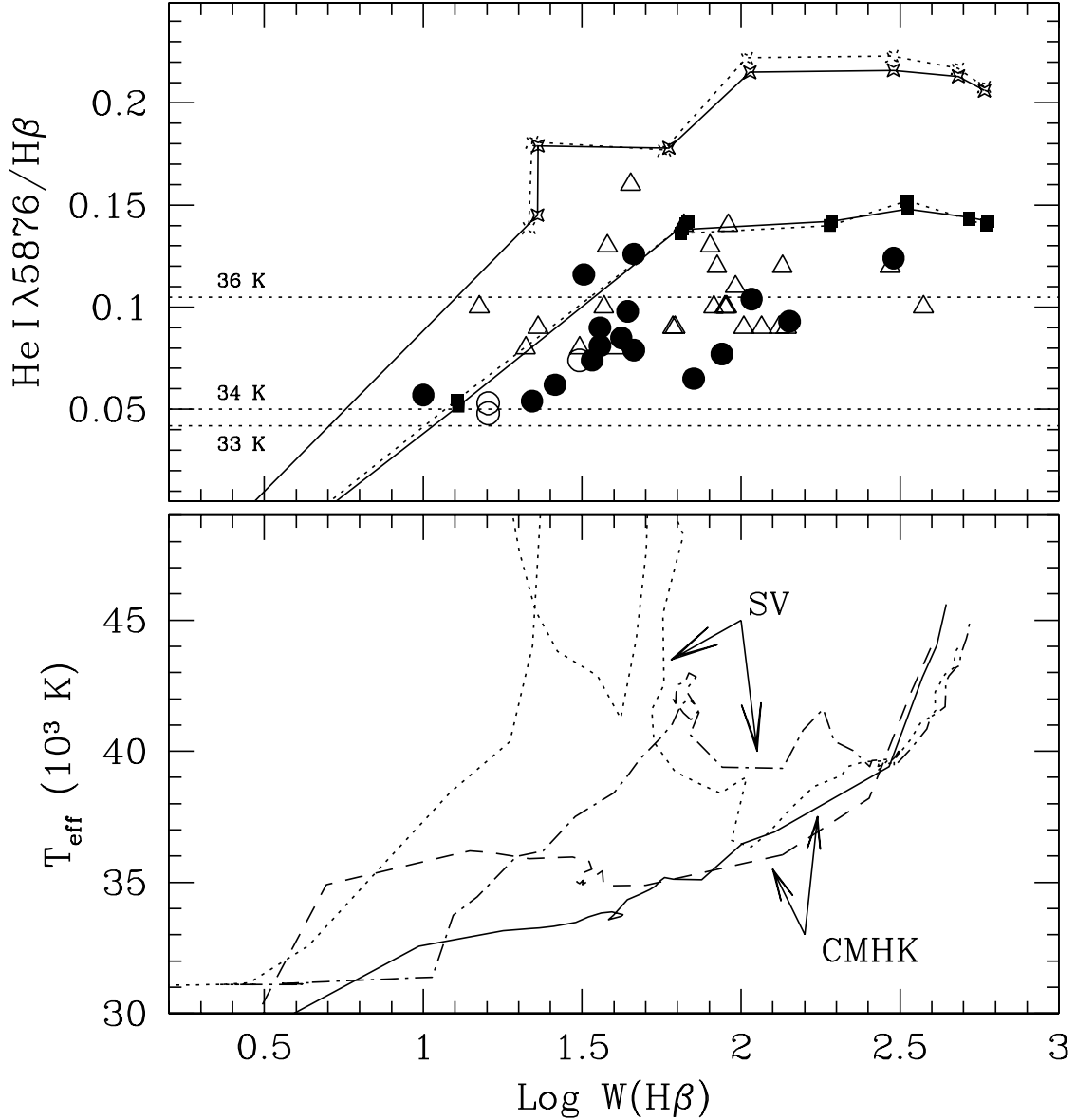}
\caption{{\em Top:} Observed He\,I\,\w5876/$H\beta$ line ratio for the
current sample (circles) and the metal-rich objects of the BKG sample
(triangles) compared to the SSL photoionization models. The latter are
shown for solar (full squares) and twice solar abundance (open stars),
and corresponding to two different cluster masses: $10^3$~\msun (full
line) and $10^6$~\msun (dotted line). The horizontal lines mark the
He\,I\,\w5876/$H\beta$ ratio corresponding to three different \teff\/
values, based on empirical results from the Galactic \hii\/ region
sample of \citet{kennicutt00}.  {\em Bottom:} Theoretical dependence
of \teff\/ on \ew, from the \citet{cervino01} models (CMHK) at two
different metal abundances (\zsun: full line; 2\,\zsun: dashed line),
and from the SV98 models (SV) (\zsun: dash-dotted line; 2\,\zsun:
dotted line). For the latter models \teff\/ was derived from the
$Q_{He}/Q_H$ ratio and the CoStar atmospheres.
\label{duo}}
\end{figure}

The He\,I\,\w5876/$H\beta$ vs.~\ew\/ plot is shown in Fig.~\ref{duo}
(top panel). Only the metal-rich objects are plotted from the BKG
sample, as estimated from the adopted abundance empirical
calibration. The results of the photoionization models of SSL, based
on the SV98 spectral energy distributions, at two different
metallicities (\zsun\/ and 2\,\zsun) are also shown (Salpeter IMF,
$M_{up}$\,=\,120~\msun). The He\,I\,\w5876/$H\beta$ ratio is independent
of the ionization parameter, as seen by comparing the models
calculated for a $10^3~$\msun\/ cluster (full lines) with those for a
$10^6$~\msun\/ cluster (dotted lines). Changes in the other parameters
explored by the SSL models have also little effect on the predicted
ratio. The predicted fluxes from the newer photoionization models are
in good agreement with the BKG models, which, as mentioned above, were
calculated based on the \citet{leitherer95} SEDs, and were therefore
dependent on different stellar tracks and stellar atmospheres than
those in the SSL work, except during the W-R phase, for which the same
\citet{schmutz92} models were employed. Empirically we find an
indication of a slight decrease of the He\,I\,\w5876/$H\beta$ ratio
with \ew, which can be interpreted as a trend of decreasing \teff\/
from $\sim$\,36,000~$K$ down to $\sim$\,34,000~$K$ as the \hii\/
regions become older. The photoionization models suggest the same
trend, but seem to predict $\sim$\,30\% stronger He\,I lines than
observed at a given age, or equivalently higher effective
temperatures. SSL, too, could not reconcile their models
with existing measurements made in samples of \hii\/ galaxies, in
particular finding no evidence for an age dependence of
He\,I\,\w5876/$H\beta$. Although such a trend appears to be present in
our current sample of EHRs, a discrepancy with the models persists.

A possible explanation for this discrepancy is that the evolutionary
models overpredict the number of He\,I ionizing photons.  In the bottom
panel of Fig.~\ref{duo} we have estimated the equivalent
\teff\/ derived from the SV98 models ($M_{up}=120~M_\odot$, Salpeter
slope), by assigning to the ionizing clusters the effective
temperature of a CoStar model atmosphere (\citealt{schaerer97}) having
the same $Q_{He}/Q_H$ ratio. The time evolution of \teff\/ thus
determined is shown by the dot-dashed (\zsun\/ models) and the dotted
(2\,\zsun) lines. It should be noted that CoStar atmospheres of solar
metallicity were used in both cases to estimate the effective
temperatures. The empirical effective temperatures from the He\,I line
are cooler than the temperatures from the models, which appear to be
boosted during the W-R phase (especially during the
WC-dominated period), which lasts for several Myr at high metallicity,
and which provides a substantial contribution to the total ionizing flux.
We also show in the bottom panel the \teff\/ trend as a function of
\ew\/ predicted by the \citet{cervino01} models
(http://www.laeff.esa.es/\~{}mcs/model) with the same upper IMF
parameters as for the SV98 case (full line:
\zsun; dashed line: 2\,\zsun). These models, an update of those
presented by \citet[see also \citealt{mashesse99}]{cervino94} use
different input parameters and assumptions regarding the stellar
atmospheres when compared with SV98 (see Table~6). The most important
differences are in the use of Geneva tracks with standard mass loss,
and in the description of the ionizing output of W-R stars, which are
treated as main sequence O stars of a given \teff, up to the O3 type,
which leads to softer ionizing spectra then the \citet{schmutz92}
continuum energy distributions. Although a different recipe was used
to derive
\teff\/ from $Q_{He}/Q_H$, based on a combination of \citet{mihalas72} and
\citet{kurucz79} atmospheres, the correspondence between \teff\/ and $Q_{He}/Q_H$
is similar to what one obtains using CoStar atmospheres. The origin of
the difference between the two sets of models in Fig.~\ref{duo},
therefore, lies mostly in the ionizing fluxes predicted by the adopted
atmosphere models.  More recent calculations by
\citet{cervino01b} include the same treatment of stellar tracks and atmospheres
as SV98, and lead to results which are consistent with the latter
work, in particular higher \teff\/'s than in the models with `soft'
W-R spectra are predicted. The comparison in Fig.~\ref{duo} shows that
the older evolutionary models, although less sofisticated in the
treatment of W-R stars, are more in agreement with the empirical
results, being able to reproduce the mild
\teff\/ gradient deduced from the He\,I line and the moderate
effective temperatures throughout most of the diagram.  We finally
note that the presence of W-R stars does not affect significantly the
derived nebular effective temperatures, since EHRs showing W-R
features in their spectra do not show signs of harder ionizing fluxes.

In conclusion, the most recent evolutionary models using a Salpeter
IMF extending to high masses appear to be consistent with several
observed line ratios, contrary to the conclusion reached by BKG on the
basis of older models. We still find some evidence, however, for an
ionizing spectrum which is possibly too hard during the W-R phase. It
is likely that the inclusion of blanketing in the atmospheres of these
stars will help resolve this issue. The recent W-R star non-LTE
blanketed model atmospheres of \citet{smith02} indeed predict a
negligible He\,II ionizing continuum flux, and a reduced He\,I
ionizing flux at high metallicities. These softer spectra, at least
qualitatively, agree with our experimental result.  On the other hand,
the unblanketed W-R models used in most current population synthesis
models have been shown to give far-UV fluxes in reasonable agreement
with observations of single W-R stars surrounded by ring nebulae,
except for the lowest-\teff\/ WNLs (\citealt{esteban93}). If, as
suggested by this latter work, such models are approximately correct
for WNE and WC stars, we would expect better agreement in our
comparison during the evolutionary phases dominated by these stars,
i.e.~around 4 Myr after the initial burst of star
formation. Fig~\ref{duo} seems to suggest that this is not the case,
and, as mentioned above, during this phase the models predict the
highest effective temperatures, which is not corroborated by the
empirical data. In this respect, the effects of the choice of
mass-loss rates adopted in the stellar evolution models on the output
of ionizing flux may become important. The number of hot W-R stars
determined by the stellar tracks can also be a matter of discussion.
The detailed investigation of the impact of all these different
scenarios is however beyond the scope of our work.

\section{Summary and conclusions}
The interpretation of the optical spectra of metal-rich \hii\/ regions
is, generally speaking, riddled with several uncertainties, owing to
the low excitation of the gas. An ill-defined electron temperature of
the nebular gas, possibly substantial contributions of underlying
stellar populations and the effects of dust can severely undermine our
ability to infer the properties of the ionizing clusters, as well as
the determination of metal abundances. We have presented a sample of
extragalactic \hii\/ regions which are estimated to have high
abundances (equal to solar and above) from semi-empirical methods
applied to strong spectral lines. Among these are a number of nuclear
hot spots, which are shown to share most of their spectral properties
with the \hii\/ regions in the disks of spiral galaxies. By using
different methods (\ew, Balmer lines in absorption, UV spectra) we
derive constistently young ages for these hot spots.

In the spectra of a number of objects in our sample we have detected
the blue bump at 4650~\AA\/ from W-R stars, and possibly also detected
WC features in the red, which allows us to put tighter constraints on
their massive stellar content. We have compared the measured intensity
and equivalent width of the 4650~\AA\/ bump with some recent
evolutionary models. By assuming instantaneous bursts of star
formation and Salpeter IMFs the ages derived from \ew\/ are in general
agreement with the predicted age span of the W-R rich phase, between 3
and 6 Myr. Some inconsistency with the models of SV98 is found
relative to the strength of the blue bump. Our favored explanation is
related to the small number of stars formed in the
clusters. Accounting for stochastic variations in the observables
appears sufficient to solve the apparent discrepancy with the
analytical models, without the need for changes in the IMF or in the
mode of star formation We await the availability of higher S/N spectra
in a larger sample of metal-rich extragalactic \hii\/ regions before
we can reach definite conclusions on the validity of this suggestion.

Some of the indicators of the hardness of the ionizing spectra
available to us, in particular the intensity of the He\,I\,\w5876
recombination line relative to $H\beta$, are examined, and compared to
the results of photoionization models by SSL. We have found that the
equivalent effective temperature of the ionizing clusters is not
affected by the presence of W-R stars. We also find some evidence that
the evolutionary models can be overestimating the number of He
ionizing photons. The observed trend of \teff\/ with \ew\/ (taken as a
good age estimator) and metal abundance are however consistent with
the predictions of \citet{cervino01}. 

Contrary to our previous conclusion from the analysis of the spectra
of extragalactic \hii\/ regions, based on evolutionary models
calculated from stellar tracks older than the ones used in the current
work, we find no compelling evidence for a depletion of massive stars
in the mass function of metal-rich clusters. The change in the models
which has mostly affected our conclusion, when compared to our
previous analysis (BKG), comes from updated stellar evolution
tracks. The non-rotating, enhanced mass loss Geneva tracks which are
part of the SV98, Starburst99 and P\'{E}GASE evolutionary models lead
to shorter W-R phases of enhanced hardness of the ionizing radiation
at high metallicity. As a result the spectra of EHRs are compatible
with the models for longer periods of time, although during the
(shorter) W-R phase the predicted ionizing flux may still be too hard.
The improvements in stellar atmospheres for massive O stars with
respect to the older models (e.g.~\/CoStar versus Kurucz) have a
secondary effect on the model emission lines, while a larger effect
might be expected from the inclusion of blanketed W-R atmospheres.

The new synthesis models, when coupled with modern photoionization
codes, seem to provide a good description of the general properties of
mostly metal-poor extragalactic \hii\/ regions, provided that
additional effects are accounted for (underlying older populations,
gas and stars affected by different amounts of extinction, loss of
ionizing photons, etc.).  This gives us some confidence that a similar
degree of accuracy can be reached by the models also in the high
metallicity regime, although, again, some issues still need further
investigation (e.g.~\/standard vs.~enhanced mass loss, W-R ionizing
fluxes). Regarding the upper IMF at high metallicity, the constraints
on the lowest value of the upper mass limit which we can still regard
as consistent with the different observables cannot be very tight,
because of the degeneracies involved (aging of the ionizing clusters,
mass function at birth) and the stochastic nature of the IMF
itself. However, when different indications from several parameters
(\ew, \teff, W-R lines, UV spectra) are analyzed in a consistent way
we conclude that the upper mass limit $M_{up}$ must be at least
40-50~\msun\/ also at metallicities well above the solar value, in
agreement with the findings of \citet{schaerer00}. Higher values of
$M_{up}$ cannot be neither ruled out nor confirmed with the presently
available data, but a mass function depleted in massive stars is not
required from modern data and evolutionary models.  The discovery and
investigation of `normal' \hii\/ regions, younger ($t<3$ Myr) than the
ones observed so far at high abundance will be crucial to settle this
question.  This might be made difficult by a possible obscuration of
young, massive stars by parental molecular clouds in metal-rich
environments.  In any case, this conclusion does not imply that
\hii\/ regions in extreme environments, such as those found in
IR-luminous starburst, might not develop peculiar initial mass
functions.

\acknowledgments
The contructive comments from the referee, C. Leitherer, helped us to
improve this work. We are grateful to him for drawing our attention to
the influence of mass-loss rate on the stellar flux.  We thank
D. Schaerer for making available to us some unpublished evolutionary
models, and M. Cervi\~{n}o for helpful discussions and for making his
newer models promptly available. FB would like to thank F. Patat for
friendly assistance at the telescope and at the ESO cafeteria.

\clearpage

\epsscale{0.5}

\epsscale{0.8}

\clearpage

\clearpage

\clearpage

\clearpage
\begin{deluxetable}{lrrl}
\tabletypesize{\scriptsize}
\tablecolumns{4}
\tablewidth{0pt}
\tablecaption{H\,II region positions and cross identification}
\tablehead{
\colhead{ID\phantom{aabbccddeeffgg}} &
\multicolumn{2}{c}{\phantom{aabb}Offset from center (arcsec)}     &
\colhead{\phantom{aabbccddee}Other ID} \\
\colhead{}              &
\colhead{\phantom{aabbcc}RA}            &
\colhead{\phantom{aabbccddee}DEC}           &
\colhead{}}
\startdata
\sidehead{M83} 
1     & \phantom{aabb}-129 & -3 &  \phantom{aabbccddee}T  230 \\
2     & \phantom{aabb}-108 & -63 &  \phantom{aabbccddee}deV  10    \\
3     & \phantom{aabb}-105 & 15 &  \phantom{aabbccddee}deV  12, D  III, W  9    \\
4     & \phantom{aabb}-98 & 28 &  \phantom{aabbccddee}deV  16    \\
5     & \phantom{aabb}-79 & -68   &  \phantom{aabbccddee}deV  22, W  8  \\
6     & \phantom{aabb}-48 & 1 &     \phantom{aabbccddee}T  172  \\
7     & \phantom{aabb}-10 & -21 &  \phantom{aabbccddee}deV  31   \\
8     & \phantom{aabb}8 & 32 &    \phantom{aabbccddee}deV  35, D  I  \\
9     & \phantom{aabb}99 & -15 &    \phantom{aabbccddee}deV  49  \\
10    & \phantom{aabb}100 & -5 &  \phantom{aabbccddee}  \\
11    & \phantom{aabb}108 & -12 &  \phantom{aabbccddee}deV  52, W  11  \\
A     & \phantom{aabb}-5.1 & -0.2 &   \phantom{aabbccddee}E  8, W  2\\
B     & \phantom{aabb}-6.0 & -4.1 &   \phantom{aabbccddee}E  4\\
C     & \phantom{aabb}-10.1 & 4.3 &   \phantom{aabbccddee}E  6, W  5 \\
D     & \phantom{aabb}-6.9 & -2.6 &    \phantom{aabbccddee}E  5, W  1\\
E     & \phantom{aabb}-2.4 & -2.2 &  \phantom{aabbccddee}  \\
\sidehead{NGC 3351}
1   & \phantom{aabb}22 & 66 &     \phantom{aabbccddee}H  34   \\
2   & \phantom{aabb}47 & -14 &     \phantom{aabbccddee}H  19   \\
A   & \phantom{aabb}-2.6 & 2.6 &     \phantom{aabbccddee}P  R2, C  K\\
B   & \phantom{aabb}-1.5 & -6.5 &    \phantom{aabbccddee}P  R3, C  J\\
C   & \phantom{aabb}4.8 & 3.6 &    \phantom{aabbccddee}P  R7, C  B\\
D   & \phantom{aabb}0.4 & 6.5 &     \phantom{aabbccddee}P  R1\\
\sidehead{NGC 6384}
1   & \phantom{aabb}-19 & 85 &    \phantom{aabbccddee}F  43   \\
2   & \phantom{aabb}23 & -71 &    \phantom{aabbccddee}F  3 \\
3   & \phantom{aabb}-3 & -87 &    \phantom{aabbccddee}F  1 \\
\enddata
\tablecomments{References for additional ID: T~=~Talbot et al.~1983;
deV~=~deVaucouleurs et al.~1983; D~=~Dufour et al.~1980;
H~=~Hodge 1974; F~=~Feinstein 1997; E~=~Elmegreen et al.~1998;
C~=~Colina et al.~1997; P~=~Planesas et al.~1997; W~=~Webster \& Smith 1983.}
\end{deluxetable}

\clearpage
\begin{deluxetable}{lcccc}
\tabletypesize{\scriptsize}
\tablecolumns{5}
\tablewidth{0pt}
\tablecaption{Equivalent widths of metallic lines in hot spots (\AA)}
\tablehead{
\colhead{ID}		& 
\colhead{Ca K}          &
\colhead{CN}		&
\colhead{G band}        &
\colhead{Mg}            \\
\colhead{(1)}   &
\colhead{(2)}   &
\colhead{(3)}   &
\colhead{(4)}   &
\colhead{(5)}  }
\startdata 
\sidehead{M83} 
A & 2.3 & 1.0 & 1.0 & 2.1 \\
B & 2.2 & 0.9 & 0.8 & 1.7 \\
C & 3.1 & 1.2 & 1.2 & 1.6 \\
D & 2.2 & 0.9 & 0.6 & 1.5 \\
E & 3.1 & 1.2 & 1.5 & 2.1 \\
\sidehead{NGC 3351}
A & 4.5 & 1.2 & 2.0 & 2.8 \\
B & 2.8 & 1.0 & 1.3 & 1.7 \\
C & 3.2 & 1.2 & 1.2 & 2.0 \\
D & 4.9 & 1.8 & 1.8 & 1.8 \\
\enddata
\end{deluxetable}

\clearpage
\begin{deluxetable}{lccccccccccccccc}
\tabletypesize{\scriptsize}
\tablecolumns{15}
\tablewidth{0pt}
\tablecaption{Line fluxes, equivalent widths and abundances}

\tablehead{
\colhead{ID}            &
\colhead{$E(B-V)$}                     &
\colhead{$W_{abs}$}     &
\colhead{[O\,II]}                &
\colhead{[O\,III]}               &
\colhead{He\,I}                  &
\colhead{[O\,I]}           &
\colhead{[N\,II]}                &
\colhead{He\,I}                  &
\colhead{[S\,II]}                &
\colhead{[S\,II]}        &
\colhead{[Ar\,III]}              &
\colhead{$W(H\beta)$}      &
\multicolumn{2}{c}{12\,+\,log(O/H)}       \\
\colhead{}                      &
\colhead{}                      &
\colhead{(\AA)}                      &
\colhead{3727}                  &
\colhead{5007}                  &
\colhead{5876}                  &
\colhead{6300}          &
\colhead{6583}          &
\colhead{6678}                  &
\colhead{6716}          &
\colhead{6731}          &
\colhead{7135}                  &
\colhead{(\AA)}          &
\colhead{MG}   &
\colhead{E}          \\
\colhead{(1)}   &
\colhead{(2)}   &
\colhead{(3)}   &
\colhead{(4)}   &
\colhead{(5)}   &
\colhead{(6)}   &
\colhead{(7)}   &
\colhead{(8)}   &
\colhead{(9)}   &
\colhead{(10)}  &
\colhead{(11)}  &
\colhead{(12)}  &
\colhead{(13)}  &
\colhead{(14)}  &
\colhead{(15)} }
\startdata
\vspace{-3mm}
\\
\sidehead{M83} 
1     & 0.28 & 4.3 &   133.2 &    31.2 &     9.8 &     \nodata &   125.8 &
\phantom{$<$}3.6 &    37.4 &    27.9 &     \nodata &     44 & 8.99 & 9.02 \\
2     & 0.32 & 0.2 &    91.4 &    10.8 &     8.1 &     3.1 &   106.7 &
\phantom{$<$}2.9 &    41.9 &    28.9 &     \nodata &     36 & 9.06 & 9.14 \\
3     & 0.33 & 1.2 &    71.2 &     5.0 &     6.5 &     1.6 &   101.6 &
\phantom{$<$}1.8 &    34.4 &    24.7 &     1.3 &     71 & 9.09 & 9.21 \\
4     & 0.43 & 1.3 &   111.0 &    18.5 &     9.3 &     2.4 &   145.8 &
\phantom{$<$}2.7 &    30.6 &    24.3 &     3.8 &    142 & 9.03 & 9.08\\
5     & 0.26 & 0 &    78.8 &    10.0 &     7.4 &     \nodata &   112.2 &
\phantom{$<$}2.4 &    31.1 &    22.6 &     1.8 &     34 & 9.07 & 9.17 \\
6     & 0.08 & 0 &    63.3 &     8.6 &     8.5 &    13.1 &   122.0 &
$<$0.6 &    32.5 &    21.9 &     \nodata &     42 & 9.09 & 9.22 \\
7     & 0.83 & 1.1 &    47.6 &    13.3 &     6.2 &     3.0 &   103.0 &
\phantom{$<$}2.4 &    32.9 &    26.4 &     0.5 &     26 & 9.09 & 9.25 \\
8     & 0.65 & 0 &    54.4 &     8.3 &     7.9 &     1.5 &   118.4 &
\phantom{$<$}2.5 &    24.8 &    19.9 &     1.5 &     46 & 9.10 & 9.25 \\
9     & 0.61 & 0 &    90.5 &     9.9 &     7.7 &     1.8 &   115.7 &
\phantom{$<$}1.7 &    31.1 &    22.8 &     2.2 &     87 & 9.06 & 9.14 \\
10    & 0.40 & 0 &   182.6 &    30.7 &    10.4 &    11.3 &   135.0 &
$<$1.9 &    66.0 &    45.6 &     \nodata &    108 & 8.94 & 8.96 \\
11    & 0.37 & 2.2 &   144.8 &    50.0 &    12.4 &     \nodata &   121.2 &
\phantom{$<$}2.5 &    34.6 &    27.6 &     5.9 &    302 & 8.96 & 8.97 \\
A     & 0.41 & 0.6 &    59.7 &     8.6 &     7.4 &     1.5 &   159.9 &
\phantom{$<$}2.6 &    22.7 &    24.7 &     1.8 &     31 & 9.09 & 9.23 \\
B     & 0.25 & 1.7 &    93.8 &    19.2 &     \nodata &     4.5 &
184.9 &     $<$0.6 &    41.6 &    44.0 &     \nodata &      7 & 9.04 & 9.11 \\
C     & 0.71 & 1.4 &    71.6 &    11.8 &     \nodata &     4.1 &
153.3 &     $<$0.2 &    34.8 &    32.7 &     1.3 &     10 & 9.08 & 9.18 \\
D     & 0.33 & 1.2 &    81.5 &    20.5 &     \nodata &     3.5 &
158.5 &     $<$0.4 &    37.7 &    38.1 &     \nodata &      9 & 9.06 & 9.13 \\
E     & 0.37 & 1.1 &   116.9 &    16.3 &     \nodata &     4.9 &
152.4 &     $<$1.4 &    49.6 &    43.1 &     \nodata &      6 & 9.03 & 9.07 \\
\sidehead{NGC 3351}
1   & 0.33 & 0.9 &    38.9 &     7.4 &     5.7 &     \nodata &    72.5 &
$<$0.3 &    21.0 &    15.2 &     \nodata &     10 & 9.11 & 9.32 \\
2   & 0.24 & 0.7 &    34.4 &     5.8 &     5.4 &     \nodata &   110.2 &
$<$0.7 &    22.9 &    15.5 &     \nodata &     22 & 9.12 & 9.36 \\
A   & 0.32 & 1.4 &    49.1 &    12.2 &     \nodata &     \nodata &   121.6
&     $<$1.5 &    30.4 &    26.9 &     \nodata &      8 & 9.10 & 9.25 \\
B   & 0.34 & 1.7 &    37.2 &     6.9 &     4.8 &     \nodata &   124.4 &
\phantom{$<$}2.5 &    25.4 &    24.7 &     1.5 &     16 & 9.11 & 9.33 \\
C   & 0.28 & 1.8 &    29.7 &     8.2 &     \nodata &     \nodata &    96.1
&     $<$1.1 &    26.1 &    23.2 &     \nodata &     12 & 9.12 & 9.37 \\
D   & 0.73 & 1.7 &    43.6 &     9.1 &     5.3 &     \nodata &   129.3 &
\phantom{$<$}3.8 &    29.5 &    26.7 &     \nodata &     16 & 9.10 & 9.29 \\
\sidehead{NGC 6384}
1   & 0.08 & 3.7 &   207.5 &    63.4 &    12.6 &     \nodata &    88.8 &
$<$3.0 &    50.3 &    36.4 &     \nodata &     46 & 8.87 & 8.89 \\
2   & 0.12 & 4.4 &   171.1 &    33.6 &    11.6 &     \nodata &   106.5 &
\phantom{$<$}2.5 &    49.7 &    32.3 &     \nodata &     32 & 8.95 & 8.96 \\
3   & 0.38 & 3.2 &   210.2 &    33.7 &     9.0 &     4.2 &   103.9 &
\phantom{$<$}3.2 &     \nodata &     \nodata &     \nodata &     36 & 8.90 & 8.92 \\
\enddata
\end{deluxetable}

\label{lines.table}
\clearpage
\begin{deluxetable}{ccc}
\tabletypesize{\scriptsize}
\tablecolumns{3}
\tablewidth{0pt}
\tablecaption{Semi-empirical abundances from model grid}
\tablehead{
\colhead{Age (Myr)} &
\colhead{12\,+\,log(O/H)}    &
\colhead{$Z/Z_\odot$}}
\startdata
\sidehead{Model: \zsun}
0 & 8.86 & 0.91 \\
1 & 8.93 & 1.06 \\
2 & 9.08 & 1.51 \\
3 & 8.93 & 1.06 \\
4 & 8.88 & 0.95 \\
5 & 8.93 & 1.07 \\
6 & 9.12 & 1.65 \\
\sidehead{Model: 2\,\zsun}
0 & 9.16 & 1.84 \\
1 & 9.26 & 2.29 \\
2 & 9.36 & 2.87 \\
3 & 9.38 & 3.04 \\
4 & 9.06 & 1.44 \\
5 & 9.10 & 1.57 \\
6 & 8.94 & 1.10 \\
\enddata
\end{deluxetable}

\clearpage
\begin{deluxetable}{ccc}
\tabletypesize{\scriptsize}
\tablecolumns{3}
\tablewidth{0pt}
\tablecaption{W-R bump: intensities and equivalent widths}
\tablehead{
\colhead{Object} &
\colhead{I(4650)/I($H\beta$)}    &
\colhead{EW(4650)}}
\startdata
M83-2 & 0.14 & 5.0 \\
M83-3 & 0.15 & 8.0 \\
M83-5 & 0.27 & 6.0 \\
M83-9 & 0.27 & 20.0 \\
M83\,A & 0.17 & 4.6 \\
\sidehead{Additional objects}
M31-4 & 0.10 & 11.0 \\
M31-5 & 0.09 & 16.0 \\
M51-4 & 0.30 & 11.0 \\
M51-6 & 0.25 & 12.0 \\
NGC 3351-4 & 0.20 & 17.0 \\
NGC 3368-1 & 0.23 & 7.0 \\
\enddata
\end{deluxetable}

\clearpage
\begin{deluxetable}{lllc}
\tabletypesize{\scriptsize}
\tablecolumns{4}
\tablewidth{0pt}
\tablecaption{Summary of evolutionary models}
\tablehead{
\colhead{Authors} &
\colhead{Stellar} &
\colhead{Stellar} &
\colhead{Mass-loss} \\
\colhead{}              &
\colhead{atmospheres}	&
\colhead{tracks}	&
\colhead{rate}}
\startdata
Cervi\~{n}o et al. (2002)	& O stars: Schaerer \& de Koter (1997) 	& Geneva & 1-2\,X\\
				& Kurucz (1992) 			& & \\
				& W-R: Schmutz et al. (1992) 		& & \\

\medskip \\

Cervi\~{n}o, Mas-Hesse \& Kunth (2001) 	& Mihalas (1972)\,+\,Kurucz (1979) & Geneva & 1\,X			\\
(Mas-Hesse \& Kunth 1991)		& W-R: Main Sequence star 	   & &			\\

\medskip \\

Leitherer et al. (1999)		& Lejeune et al. (1997)			& Geneva & 1-2\,X \\
{\em (Starburst99)}		& W-R: Schmutz et al. (1992)		& & \\

\medskip \\

Schaerer \& Vacca (1998)	& O stars: Schaerer \& de Koter (1997) 	& Geneva & 2\,X \\
				& Kurucz (1992) 			& & \\
				& W-R: Schmutz et al. (1992) 		& & \\

\medskip \\

Fioc \& Rocca-Volmerange (1997)		& Lejeune et al. (1997)		& Padova & 1\,X 		\\
{\em (P\'{E}GASE 2)}			& $T>50,000~K$: Clegg \& Middlemass (1987) & &	\\
\enddata
\end{deluxetable}


\clearpage

\begin{thebibliography}{}

\bibitem[Achtermann \& Lacy\/(1995)]{achtermann95} Achtermann, J.M. \&
Lacy, J.H. 1995, \apj, 439, 163

\bibitem[Alloin \& Nieto\/(1982)]{alloin82} Alloin, D. \& Nieto,
J.-L. 1982, \aaps, 50, 491

\bibitem[Alloin et al.\/(1979)]{alloin79} Alloin, D., Collin-Souffrin,
S., Joly, M. \& Vigroux, L. 1979, \aap, 78, 200

\bibitem[Alonso-Herrero, Ryder \& Knapen\/(2001)]{herrero01} Alonso-Herrero, A.,
Ryder, S.D. \& Knapen, J.H. 2001, \mnras, 322, 757

\bibitem[Arnault, Kunth \& Schild\/(1989)]{arnault89} Arnault, P.,
Kunth, D. \& Schild, H. 1989, \aap, 224, 73

\bibitem[Barth \& Shields\/(2000)]{barth00} Barth, A.J., \& Shields,
J.C. 2000, \pasp, 112, 753

\bibitem[Beck, Kelly \& Lacy\/(1997)]{beck97} Beck, S.C., Kelly, D.M. \& Lacy,
J.H. 1997, \aj, 114, 585

\bibitem[Bernasconi \& Maeder\/(1996)]{bernasconi96} Bernasconi,
P.A. \& Maeder, A. 1996, \aap, 307, 829

\bibitem[Bica\/(1988)]{bica88} Bica, E. 1988, \aap, 195, 76

\bibitem[Bica \& Alloin\/(1987)]{bica87} Bica, E. \& Alloin, D. 1987,
\aaps, 70, 281

\bibitem[Bohlin et al.\/(1983)]{bohlin83} Bohlin, R.C., Cornett, R.H.,
Hill, J.K., Smith, A.M. \& Stecher, T.P. 1983, \apj, 274, L53

\bibitem[B\"{o}ker et al.\/(2000)]{boker00} B\"{o}ker, T., van der Marel,
R.P., Mazzuca, L., Rix, H.-W., Rudnick, G., Ho, L.C. \& Shields,
J.C., 2001, \aj, 121, 1473

\bibitem[Bresolin, Kennicutt \& Garnett\/(1999)]{bresolin99} Bresolin,
F., Kennicutt, R.C. \& Garnett, D.R. 1999, \apj, 510, 104

\bibitem[Buta \& Crocker\/(1993)]{buta93} Buta, R. \& Crocker,
D.A. 1993, \aj, 105, 1344

\bibitem[Calzetti\/(2001)]{calzetti01} Calzetti, D. 2001, \pasp, 113, 1449

\bibitem[Calzetti, Kinney \& Storchi-Bergmann\/(1994)]{calzetti94}
Calzetti, D., Kinney, A.L. \& Storchi-Bergmann, T. 1994, \apj, 429, 582

\bibitem[Castellanos, D\'{\i}az \& Terlevich\/(2001)]{castellanos01} 
Castellanos, M., D\'{\i}az, A.I. \& Terlevich, E. 2001, MNRAS, in
press

\bibitem[Cervi\~{n}o et al.\/(2002)]{cervino01b} Cervi\~{n}o, M.,
Valls-Gabaud, D., Luridiana, V. \& Mas-Hesse, J.M. 2002, \aap, 381, 51

\bibitem[Cervi\~{n}o, Mas-Hesse \& Kunth\/(2001)]{cervino01} Cervi\~{n}o, M.,
Mas-Hesse, J.M. \& Kunth, D. 2001, \aap, submitted

\bibitem[Cervi\~{n}o, Luridiana \& Castander\/(2000)]{cervino00} Cervi\~{n}o,
M., Luridiana, V. \& Castander, F.J. 2000, \aap, 360, L5

\bibitem[Cervi\~{n}o \& Mas-Hesse\/(1994)]{cervino94} Cervi\~{n}o, M., \&
Mas-Hesse, J. M. 1994, \aap, 284, 749

\bibitem[Clegg \& Middlemass\/(1987)]{clegg87} Clegg, R.E.S. \&
Middlemass, D. 1987, \mnras, 228, 759

\bibitem[Colina et al.\/(1997)]{colina97} Colina, L., Garc\'{\i}a Vargas,
M.L., Mas-Hesse, J.M., Alberdi, A. \& Krabbe, A. 1997, \apj, 484, L41

\bibitem[Conselice et al.\/(2000)]{conselice00} Conselice, C.J.,
Gallagher, J.S., Calzetti, D., Homeier, N. \& Kinney, A. 2000, \aj,
119, 79

\bibitem[Conti\/(1991)]{conti91} Conti, P.S. 1991, \apj, 377, 115

\bibitem[Contini et al.\/(2002)]{contini01} Contini, T., Treyer, M.A.,
Sullivan, M. \& Ellis, R.S. 2002, \mnras, 330, 75

\bibitem[Copetti, Pastoriza \& Dottori\/(1986)]{copetti86} Copetti,
M.V.F., Pastoriza, M.G. \& Dottori, H.A. 1986, \aap, 156, 111

\bibitem[Coziol, Doyon \& Demers\/(2001)]{coziol01} Coziol, R., Doyon,
R. \& Demers, S.  2001, \mnras, 325, 1081

\bibitem[Crowther\/(1999)]{crowther99b} Crowther, P.A. 1999, in IAU
Symposium 193, Wolf-Rayet Phenomena in Massive Stars and Starburst
Galaxies, ed. K.A. van der Hucht, G. Koenigsberger \& P.R.J. Eenens
(San Francisco: ASP), 129

\bibitem[Crowther et al.\/(1999)]{crowther99} Crowther, P.A., Pasquali,
A., de Marco, O., Schmutz, W., Hillier, D.J., \& de Koter, A., 1999,
\aap, 350, 1007

\bibitem[de Jager, Nieuwenhuijzen \& van der Hucht\/(1988)]{dejager88}
de Jager, C., Nieuwenhuijzen, H. \& van der Hucht, K.A. 1988, \aaps,
72, 259

\bibitem[de Vaucouleurs, Pence \& Davoust\/(1983)]{devau83} de
Vaucouleurs, G., Pence, W. D., \& Davoust, E. 1983, \apjs, 53, 17

\bibitem[D\'{\i}az et al.\/(2000)]{diaz00} D\'{\i}az, A.I., Castellanos, M.,
Terlevich, E. \& Garc\'{\i}a-Vargas, M.L. 2000, MNRAS, 318, 462

\bibitem[D\'{\i}az et al.\/(1991)]{diaz91} D\'{\i}az, A.I., Terlevich, E.,
V\'{\i}lchez, J.M., Pagel, B.E.J. \& Edmunds, M.G. 1991, \mnras, 253, 245

\bibitem[Doherty et al.\/(1995)]{doherty95} Doherty, R.M., Puxley,
P.J., Lumsden, S.L. \& Doyon, R. 1995, \mnras, 277, 577

\bibitem[Dopita et al.\/(2000)]{dopita00} Dopita, M.A., Kewley, L.J., Heisler, C.A. \& Sutherland, R.S. 2000, \apj, 542, 224

\bibitem[Dopita \& Evans\/(1986)]{dopita86} Dopita, M.A. \& Evans,
I.N. 1986, ApJ, 307, 431

\bibitem[Dufour et al.\/(1980)]{dufour80} Dufour, R.J., Talbot, R.J.,
Jensen, E.B. \& Shields, G.A. 1980, \apj, 236, 119

\bibitem[Dutil \& Roy\/(1999)]{dutil99} Dutil, Y. \& Roy, J.-R. 1999,
\apj, 516, 62

\bibitem[Edmunds\/(1989)]{edmunds89} Edmunds, M.G. 1989, in
Evolutionary Phenomena in Galaxies, ed. J.E. Beckman \& B.E.J. Pagel
(Cambridge: Cambridge University Press), 356

\bibitem[Edmunds \& Pagel\/(1984)]{ep84} Edmunds, M.G. \& Pagel,
B.E.J. 1984, MNRAS, 211, 507

\bibitem[Elmegreen, Chromey \& Warren\/(1998)]{elmegreen98} Elmegreen,
D. M., Chromey, F. R., \& Warren, A. R. 1998, \aj, 116, 2834

\bibitem[Elmegreen et al.\/(1997)]{elmegreen97} Elmegreen, D.M.,
Chromey, F.R., Santos, M. \& Marshall, D. 1997, \aj, 114, 1850

\bibitem[Elmegreen\/(1994)]{elmegreen94} Elmegreen, B.G. 1994, \apj,
425, L73

\bibitem[Esteban et al.\/(1993)]{esteban93} Esteban, C., Smith, L.J.,
Vilchez, J.M. \& Clegg, R.E.S. 1993, \aap, 272, 299

\bibitem[Feinstein\/(1997)]{feinstein97} Feinstein, C. 1997, \apjs, 112, 29

\bibitem[Ferland\/(1997)]{ferland97} Ferland, G.J., 1997, Hazy, a
brief introduction to Cloudy 90

\bibitem[Figer et al.\/(1998)]{figer98} Figer, D.F., Najarro, F.,
Morris, M., McLean, I.S., Geballe, T.R., Ghez, A.M. \& Langer,
N. 1998, \apj, 506, 384

\bibitem[Fioc \& Rocca-Volmerange\/(1997)]{fioc97} Fioc, M. \&
Rocca-Volmerange, B. 1997, \aap, 326, 950

\bibitem[Freedman et al.\/(2001)]{freedman01} Freedman, W.L. et
al. 2001, \apj, 553, 47

\bibitem[Gabler et al.\/(1989)]{gabler89} Gabler, R., Gabler, A.,
Kudritzki, R.P., Puls, J. \& Pauldrach, A. 1989, \aap, 226, 162

\bibitem[Gallais et al.\/(1991)]{gallais91} Gallais, P., Rouan, D.,
Lacombe, F., Tiph\`{e}ne, D. \& Vauglin, I. 1991, \aap, 243, 309

\bibitem[Garnett\/(1992)]{garnett92} Garnett, D.R. 1992, \aj, 103, 1330

\bibitem[Goldader et al.\/(1997)]{goldader97} Goldader, J.D., Joseph,
R.D., Doyon, R. \& Sanders, D.B. 1997, \apj, 474, 104

\bibitem[Gonz\'{a}lez Delgado\/(2001)]{delgado01} Gonz\'{a}lez Delgado,
R.M. 2001, in IAU Symposium 207, Extragalactic Star Clusters,
ed. E.K. Grebel, D. Geisler \& D. Minniti (San Francisco: ASP)

\bibitem[Gonz\'{a}lez Delgado, Heckman \&
Leitherer\/(2001)]{gonzalez01} Gonz\'{a}lez Delgado, R.M., Heckman,
T. \& Leitherer, C. 2001, \apj, 546, 845

\bibitem[Gonz\'{a}lez Delgado \& Per\'{e}z\/(2000)]{delgado00} Gonz\'{a}lez
Delgado, R.M. \& Per\'{e}z, E. 2000, \mnras, 317, 64

\bibitem[Gonz\'{a}lez Delgado, Leitherer \& Heckman\/(1999)]{gonzalez99}
Gonz\'{a}lez Delgado, R.M., Leitherer, C. \& Heckman, T.M. 1999,
\apjs, 125, 489

\bibitem[Gonz\'{a}lez Delgado et al.\/(1999b)]{gonzalez99c}
Gonz\'{a}lez Delgado, R.M., Garc\'{\i}a-Vargas, M.L., Goldader, J.,
Leitherer, C. \& Pasquali, A. 1999, \apj,
513, 707

\bibitem[Gonz\'{a}lez Delgado \& Leitherer\/(1999)]{gonzalez99b} Gonz\'{a}lez
Delgado, R.M. \& Leitherer, C. 1999, \apjs, 125, 479

\bibitem[Guseva, Izotov \& Thuan\/(2000)]{guseva00} Guseva, N.G.,
Izotov, Y.I. \& Thuan, T.X. 2000, \apj, 531, 776

\bibitem[Hamuy et al.\/(1992)]{hamuy92} Hamuy, M., Walker, A.R.,
Suntzeff, N.B., Gigoux, P., Heathcote, S.R., \& Phillips,
M.M. 1992, \pasp, 104, 533

\bibitem[Harris et al.\/(2001)]{harris01} Harris, J., Calzetti, D.,
Gallagher, J.S, Conselice, C.J. \& Smith, D.A. 2001, \aj, 122, 3046

\bibitem[Heap et al.\/(1993)]{heap93} Heap, S.R., Holbrook, J.,
Malumuth, E., Shore, S. \& Waller, W., 1993, BAAS, 182, 3104

\bibitem[Heckman et al.\/(1998)]{heckman98} Heckman, T.M., Robert, C.,
Leitherer, C., Garnett, D.R. \& van der Rydt, F. 1998, \apj, 503, 646

\bibitem[Ho, Filippenko \& Sargent\/(1997)]{ho97} Ho, L.C., 
Filippenko, A.V. \& Sargent, W.L.W. 1997, \apj, 487, 579

\bibitem[Hodge\/(1974)]{hodge74} Hodge, P. 1974, \apjs, 27, 113

\bibitem[Hummer \& Storey\/(1987)]{hummer87} Hummer, D.G. \& Storey,
P.J. 1987, \mnras, 224, 801

\bibitem[Israel \& Baas\/(2001)]{israel01} Israel, F.P. \& Baas,
F. 2001, \aap, 371, 433

\bibitem[Johnson et al.\/(2000)]{johnson00} Johnson, K.E., Leitherer,
C., Vacca, W.D. \& Conti, P.S. 2000, \aj, 120, 1273

\bibitem[Johnson et al.\/(1999)]{johnson99} Johnson, K.E., Vacca, W.D.,
Leitherer, C., Conti, P.S. \& Lipscy, S.J. 1999, \aj, 117, 1708

\bibitem[Kenney et al.\/(1992)]{kenney92} Kenney, J., Wilson, C.,
Scoville, N., Devereux, N. \& Young, J. 1992, \apj, 395, L79

\bibitem[Kennicutt et al.\/(2000)]{kennicutt00} Kennicutt, R. C.,
Bresolin, F., French, H. \& Martin, P. 2000, \apj, 537, 589

\bibitem[Kennicutt, Keel \& Blaha\/(1989)]{kennicutt89} Kennicutt, R.C., Keel,
W.C. \& Blaha, C.A. 1989, \aj, 97, 1022


\bibitem[Kewley et al.\/(2001)]{kewley01} Kewley, L.J., Dopita, M.A.,
Sutherland, R.S., Heisler, C.A. \& Trevena, J. 2001, \apj, 556, 121

\bibitem[Kinkel \& Rosa\/(1994)]{kinkel94} Kinkel, U. \& Rosa,
M.R. 1994, A\&A, 2822, L37

\bibitem[Kobulnicky, Kennicutt \& Pizagno\/(1999)]{kobulnicky99} 
Kobulnicky, H.A., Kennicutt, R.C. \& Pizagno, J.L. 1999, \apj, 514,
544

\bibitem[Kotilainen et al.\/(2000)]{kotilainen00} Kotilainen, J.K.,
Reunanen, J., Laine, S. \& Ryder, S.D. 2000, \aap, 353, 834

\bibitem[Kudritzki \& Puls\/(2000)]{kud00} Kudritzki, R.P. \& Puls,
J. 2000, ARAA, 38, 613

\bibitem[Kurucz\/(1992)]{kurucz92} Kurucz, R.L. 1992, in IAU Symposium 149,
The Stellar Populations of Galaxies, ed. B. Barbuy \& A. Renzini
(Dordrecht: Kluwer), 225

\bibitem[Kurucz\/(1979)]{kurucz79} Kurucz, R.L. 1979, \apjs, 40, 1

\bibitem[Leitherer\/(2001)]{leitherer02} Leitherer, C. 2001, in A
Decade of HST Observations, ed. M. Livio, K.S. Noll \& M. Stiavelli
(Cambridge: Cambridge University Press), in press

\bibitem[Leitherer et al.\/(2001)]{leitherer01} Leitherer, C., Leao,
J.R.S., Heckman, T.M., Lennon, D.J., Pettini, M. \& Robert, C. 2001,
\apj, 550, 724

\bibitem[Leitherer et al.\/(1999)]{leitherer99} Leitherer, C.,
et al. 1999, \apjs, 123, 3

\bibitem[Leitherer\/(1999)]{leitherer99iau} Leitherer, C. 1999, in
IAU Symposium 193, Wolf-Rayet Phenomena in Massive Stars and Starburst
Galaxies, ed. K.A. van der Hucht, G. Koenigsberger \& P.R.J. Eenens
(San Francisco: ASP), 526

\bibitem[Leitherer, Chapman \& Koribalski\/(1997)]{leitherer97}
Leitherer, C., Chapman, J.M. \& Koribalski, B. 1997, \apj, 481, 898

\bibitem[Leitherer \& Heckman\/(1995)]{leitherer95} Leitherer, C. \&
Heckman, T.M. 1995, \apjs, 96, 9

\bibitem[Leitherer, Robert \& Heckman\/(1995)]{leitherer95b}
Leitherer, C., Robert, C. \&
Heckman, T.M. 1995, \apjs, 99, 173

\bibitem[Lejeune, Buser \& Cuisinier\/(1997)]{lejeune97} Lejeune, T.,
Buser, R. \& Cuisinier, F. 1997, \aaps, 125, 229

\bibitem[Luhman et al.\/(1998)]{luhman98} Luhman, M.L., Satyapal, S.,
Fischer, J., Wolfire, M.G., Cox, P., Lord, S.D., Smith, H.A., Stacey,
G.J. \& Unger, S.J. 1998, \apj, 504, L11

\bibitem[Maeder \& Meynet\/(2001)]{maeder01} Maeder, A. \& Meynet,
G. 2001, \aap, 373, 555

\bibitem[Maeder \& Meynet\/(1994)]{maeder94} Maeder, A. \& Meynet,
G. 1994, A\&A, 287, 803

\bibitem[Maeder\/(1990)]{maeder90} Maeder, A. 1990, A\&AS, 84, 139

\bibitem[Maoz et al.\/(2001)]{maoz01} Maoz, D., Barth, A.J., Ho, L.C.,
Sternberg, A. \& Filippenko, A.V. 2001, \aj, 121, 3048

\bibitem[Martin \& Friedli\/(1999)]{martin99} Martin, P. \& Friedli,
D. 1999, \aap, 346, 769

\bibitem[Massey\/(1998)]{massey98} Massey, P. 1998, in The Stellar
Initial Mass Function, ed. G. Gilmore \& D. Howell (San Francisco:
ASP), 17

\bibitem[Massey \& Hunter\/(1998)]{masseyhunter98} Massey, P. \&
Hunter, D.A. 1998, \apj, 493, 180

\bibitem[Mas-Hesse \& Kunth\/(1999)]{mashesse99} Mas-Hesse, J.M. \&
Kunth, D. 1999, \aap, 349, 765

\bibitem[McCall\/(1984)]{mccall84} McCall, M. L. 1984, \mnras, 208, 253

\bibitem[McCall, Rybski \& Shields\/(1985)]{mccall85} McCall, M.L., 
Rybski, P.M. \& Shields, G.A. 1985, ApJS, 57, 1

\bibitem[McGaugh\/(1994)]{mcgaugh94} McGaugh, S.S. 1994, ApJ, 426, 135

\bibitem[McGaugh\/(1991)]{mcgaugh91} McGaugh, S.S. 1991, ApJ, 380, 140

\bibitem[Meynet \& Maeder\/(2000)]{meynet00} Meynet, G. \& Maeder,
A. 2000, \aap, 361, 101

\bibitem[Meynet et al.\/(1994)]{meynet94} Meynet, G., Maeder, A.,
Schaller, G., Schaerer, D., \& Charbonnel, C. 1994, \aap S, 103, 97

\bibitem[Meynet\/(1995)]{meynet95} Meynet, G. 1995, \aap, 298, 767

\bibitem[Mihalas\/(1972)]{mihalas72} Mihalas, D. 1972, Non-LTE model
atmospheres for B and O stars, NCAR-TN/STR-76

\bibitem[Moy, Rocca-Volmerange \& Fioc\/(2001)]{moy01} Moy, E., 
Rocca-Volmerange, B. \& Fioc, M. 2001, A\&A, 365, 347

\bibitem[Morgan\/(1958)]{morgan58} Morgan, W.W. 1958, \pasp, 70, 364

\bibitem[Nugis \& Lamers\/(2000)]{nugis00} Nugis, T. \& Lamers,
H.J.G.L.M. 2000, \aap, 360, 227

\bibitem[Oey \& Kennicutt\/(1993)]{oey93} Oey, M.S. \& Kennicutt,
R.C. 1993, \apj, 411, 137

\bibitem[Origlia et al.\/(2001)]{origlia01} Origlia, L., Leitherer,
C., Aloisi, A., Greggio, L. \& Tosi, M. 2001, \aj, 122, 815

\bibitem[Osterbrock\/(1989)]{osterbrock} Osterbrock, D.E. 1989, The
Astrophysics of Gaseous Nebulae and Active Galactic Nuclei (Mill
Valley: University Science Books)

\bibitem[Pagel et al.\/(1979)]{pagel79} Pagel, B.E.J., Edmunds, M.G.,
Blackwell, D.E., Chun, M.S. \& Smith, G. 1979, MNRAS, 189, 95

\bibitem[Panagia\/(2000)]{panagia00} Panagia, N. 2000, in The
evolution of the Milky Way: Stars Versus Clusters, ed. F. Matteucci \&
F. Giovanelli (Dordrecht: Kluwer), 495

\bibitem[P\'{e}rez-Ram\'{\i}rez et al.\/(2000)]{perez00} 
P\'{e}rez-Ram\'{\i}rez, D., Knapen, J.H., Peletier, R.F., Laine, S.,
Doyon, R. \& Nadeau, D. 2000, \mnras, 317, 234

\bibitem[Planesas, Colina \& P\'{e}rez-Olea\/(1997)]{planesas97} 
Planesas, P., Colina, L. \& P\'{e}rez-Olea, D. 1997, \aap, 325, 81

\bibitem[Puxley, Doyon \& Ward\/(1997)]{puxley97} Puxley, P.J., Doyon, R. \&
Ward, M.J. 1997, \apj, 476, 120

\bibitem[Pilyugin\/(2001b)]{pyliugin01b} Pilyugin, L.S. 2001b, A\&A, 373, 56

\bibitem[Pilyugin\/(2001a)]{pyliugin01} Pilyugin, L.S. 2001a, A\&A, 369, 594

\bibitem[Pilyugin\/(2000)]{pyliugin00} Pilyugin, L.S. 2000, A\&A, 362, 325

\bibitem[Raimann et al.\/(2000)]{raimann00} Raimann, D., Bica, E.,
Storchi-Bergmann, T., Melnick, J. \& Schmitt, H. 2000, \mnras, 314, 295

\bibitem[Rosa \& D'Odorico\/(1986)]{rosa86} Rosa, M., \& D'Odorico,
S. 1986, in IAU Symposium 116, Luminous stars and associations in
galaxies (Dordrecht: Reidel), 355

\bibitem[Rouan et al.\/(1996)]{rouan96} Rouan, D., Tiph\`{e}ne, D., Lacombe,
F., Boulade, O., Clavel, J., Gallais, P., Metcalfe, L., Pollock, A.
\& Siebenmorgen, R. 1996 \aap, 315, L141

\bibitem[Ryder, Knapen \& Takamiya\/(2001)]{ryder01} Ryder, S.D., 
Knapen, J.H. \& Takamiya, M. 2001, \mnras, 323, 663

\bibitem[Saraiva et al.\/(2001)]{saraiva01} Saraiva, M.F., Bica, E.,
Pastoriza, M.G. \& Bonatto, C. 2001, \aap, 376, 43

\bibitem[Schaerer et al.\/(2000)]{schaerer00} Schaerer, D., Guseva,
N.G., Izotov, Y.I. \& Thuan, T.X. 2000, A\&A, 362, 53

\bibitem[Schaerer, Contini \& Pindao\/(1999)]{schaerer99} Schaerer, D., Contini,
T. \& Pindao, M. 1999, \aaps, 136, 35

\bibitem[Schaerer\/(1999)]{schaerer99b} Schaerer, D. 1999, in
IAU Symposium 193, Wolf-Rayet Phenomena in Massive Stars and Starburst
Galaxies, ed. K.A. van der Hucht, G. Koenigsberger \& P.R.J. Eenens
(San Francisco: ASP), 539

\bibitem[Schaerer \& Vacca\/(1998)]{schaerer98} Schaerer, D., \& Vacca,
W. D. 1998, \apj, 497, 618

\bibitem[Schaerer \& de Koter\/(1997)]{schaerer97} Schaerer, D., \& de
Koter, A. 1997, \aap, 322, 598

\bibitem[Schaerer et al.\/(1996)]{schaerer96} Schaerer, D., de Koter,
A., Schmutz, W., \& Maeder, A. 1996, \aap, 310, 837

\bibitem[Schaller et al.\/(1992)]{schaller92} Schaller, G., Schaerer,
D., Meynet, G. \& Maeder, A. 1992, \aaps, 96, 269

\bibitem[Schlegel, Finkbeiner \& Davis\/(1998)]{schlegel98} Schlegel, D.J.,
Finkbeiner, D.P. \& Davis, M. 1998, \apj, 500. 525

\bibitem[Schmutz, Leitherer \& Gruenwald\/(1992)]{schmutz92} Schmutz,
W., Leitherer, C. \& Gruenwald, R. 1992, \pasp, 104, 1164

\bibitem[Seaton\/(1979)]{seaton79} Seaton, M.J. 1979, \mnras, 187, 73

\bibitem[Sersic \& Pastoriza\/(1967)]{sersic67} Sersic, J.L. \&
Pastoriza, M. 1967, \pasp, 79, 152

\bibitem[Sersic \& Pastoriza\/(1965)]{sersic65} Sersic, J.L. \&
Pastoriza, M. 1965, \pasp, 77, 287

\bibitem[Shields \& Kennicutt\/(1995)]{shields95} Shields, J.C. \&
Kennicutt, R.C. 1995, \apj, 454, 807

\bibitem[Skillman\/(1989)]{skillman89} Skillman, E.D. 1989, ApJ, 347, 883

\bibitem[Shlosman\/(1999)]{shlosman99} Shlosman, I. 1999, in The
Evolution of Galaxies on Cosmological Timescales, ed. J.E. Beckman \&
T.J. Mahoney (San Francisco: ASP), 100

\bibitem[Smith, Norris \& Crowther\/(2002)]{smith02} Smith, L.J.,
Norris, R.P.F. \& Crowther, P.A. 2002, \apss, in press

\bibitem[Smith, Shara \& Moffat\/(1996)]{smith96} Smith, L.F., Shara,
M.M. \& Moffat, A.F.J. 1996, \mnras, 281, 163

\bibitem[Smith\/(1968)]{smith68} Smith, L.F. 1968, \mnras, 138, 109

\bibitem[Stasi\'{n}ska, Schaerer \& Leitherer\/(2001)]{stasinska01b}
Stasi\'{n}ska, G., Schaerer, D. \& Leitherer, C. 2001, \aap, 370, 1

\bibitem[Stasi\'{n}ska\/(2001)]{stasinska01} Stasi\'{n}ska, G. 2001,
RevMexAA, in press

\bibitem[Stasi\'{n}ska\/(1999)]{stasinska99} Stasi\'{n}ska, G. 1999, in Dwarf
Galaxies and Cosmology, ed. T.X. Thuan, C. Balkowski, V. Cayette \&
J. Tran Thanh Van (Gif-sur-Yvette: Editions Frontieres), 259

\bibitem[Stasi\'{n}ska \& Leitherer\/(1996)]{stasinska96} Stasi\'{n}ska, G. \&
Leitherer, C. 1996, \apjs, 107, 661

\bibitem[Storchi-Bergmann, Kinney \& Challis\/(1995)]{storchi95} 
Storchi-Bergmann, T., Kinney, A.L. \& Challis, P. 1995, \apjs, 98, 103

\bibitem[Storchi-Bergmann, Calzetti \& Kinney\/(1994)]{storchi94} 
Storchi-Bergmann, T., Calzetti, D. \& Kinney, A.L. 1994, \apj, 429,
572

\bibitem[Telesco\/(1988)]{telesco88} Telesco, C. M. 1988, \araa, 26, 343

\bibitem[Thatte, Tecza \& Genzel\/(2000)]{thatte00} Thatte, N., Tecza,
M. \& Genzel, R., 2000, \aap, 364, L47

\bibitem[Thornley et al.\/(2000)]{thornley00} Thornely, M.D.,
F\"{o}rster Schreiber, N.M., Lutz, D., Genzel, R., Spoon, H.W.W. \&
Kunze, D. 2000, \apj, 539, 641

\bibitem[Tremonti et al.\/(2001)]{tremonti01} Tremonti, C.A., Calzetti,
D., Leitherer, C. \& Heckman, T.M. 2001, \apj, 555, 322

\bibitem[Turner, Ho \& Beck\/(1987)]{turner87} Turner, J.L., Ho,
P.T.P. \& Beck, S.C. 1987, \apj, 313, 644

\bibitem[Vacca \& Conti\/(1992)]{vacca92} Vacca, W.D. \& Conti,
P.S. 1992, \apj, 401, 543

\bibitem[van Zee et al.\/(1998)]{vanzee98} van Zee, L., Salzer, J.J.,
Haynes, M.P., O'Donoghue, A.A. \& Balonek, T.J. 1998, \aj, 116, 2805

\bibitem[Garc\'{\i}a-Vargas, Bressan \& D\'{\i}az\/(1995)]{vargas95}
Garc\'{\i}a-Vargas, M.L., Bressan, A. \& D\'{\i}az, A.I. 1995, \aaps, 112, 13

\bibitem[Veilleux \& Osterbrock\/(1987)]{veilleux87} Veilleux, S., \&
Osterbrock, D. E. 1987, \apjs, 63,295

\bibitem[Vila-Costas \& Edmunds\/(1992)]{vilacostas92} Vila-Costas,
M. B., \& Edmunds, M. G. 1992, \mnras, 259, 121

\bibitem[Webster \& Smith\/(1983)]{webster83} Webster, B.L. \& Smith,
M.G. 1983, \mnras, 204, 743

\bibitem[Zaritsky, Kennicutt \& Huchra\/(1994)]{zaritsky94} Zaritsky,
D., Kennicutt, R.C. \& Huchra, J.P. 1994, \apj, 420, 87

\bibitem[Walborn, Nichols-Bohlin \& Panek\/(1985)]{walborn85} Walborn,
N.R., Nichols-Bohlin, J. \& Panek, R.J. 1985, International
Ultraviolet Explorer Atlas of O-type Spectra from 1200 to 1900 \AA\/
(Washington: NASA)

\end{thebibliography}
\end{document}